\documentclass[12pt]{iopart}

\pdfoutput = 1
\usepackage{bm}
\usepackage{color}
\usepackage{graphicx}
\usepackage{citesort}
\usepackage{cancel}
\eqnobysec

\newcommand{\lwig}{\mbox{\;\raisebox{.3ex}
    {$<$}$\!\!\!\!\!$\raisebox{-.9ex}{$\sim$}\;}}
\newcommand{\gwig}{\mbox{\;\raisebox{.3ex}
    {$>$}$\!\!\!\!\!$\raisebox{-.9ex}{$\sim$}}\;}
\newcommand{\lambdabar}{{\hbox{$\lambda$\kern-1.ex\raise+0.45ex\hbox{--}}}}
\newcommand{\be}{\begin{equation}}
\newcommand{\ee}{\end{equation}}
\newcommand{\bea}{\begin{eqnarray}}
\newcommand{\eea}{\end{eqnarray}}

\def\eV{{\rm \ eV}}
\def\GeV{{\rm \ GeV}}

\def\keV{{\rm \ keV}}
\def\TeV{{\rm \ TeV}}

 \def\gae{\; ^{>}_{\sim} \;}

\DeclareMathAlphabet{\mathpzc}{OT1}{pzc}{m}{it}

\begin{document}

\begin{flushright}
{\large \tt 
TTK-11-34\\
ULB-TH/11-20}
\end{flushright}

\title[Asymmetric Inert Doublet DM]{Asymmetric Inelastic Inert Doublet Dark Matter from 
Triplet Scalar Leptogenesis}

\author{Chiara~Arina\dag\ and Narendra~Sahu\ddag}

\address{\dag\  
Institut f\"ur Theoretische Teilchenphysik und Kosmologie, RWTH Aachen, 52056 Aachen, Germany}
\address{\ddag\ 
Service de Physique Th\'eorique, Universit\'e Libre de Bruxelles, CP225, Bld du Triomphe, 1050 Brussels, Belgium}

\eads{\mailto{chiara.arina@physik.rwth-aachen.de}, \mailto{Narendra.Sahu@ulb.ac.be}}

\begin{abstract}
The nature of dark matter (DM) particles and the mechanism that provides their measured relic 
abundance are currently unknown. In this paper we investigate inert scalar 
and vector like fermion doublet DM candidates with a charge asymmetry in the dark sector, which is 
generated by the same mechanism that provides the baryon asymmetry, namely baryogenesis-via-leptogenesis induced by decays of scalar triplets. At the same time the model gives rise to neutrino masses in the ballpark of oscillation experiments via type II seesaw. We discuss possible sources of depletion of asymmetry in the DM and visible 
sectors and solve the relevant Boltzmann equations for quasi-equilibrium decay of triplet scalars. A 
Monte-Carlo-Markov-Chain analysis is performed for the whole parameter space. The survival of the asymmetry 
in the dark sector leads to inelastic scattering off nuclei. We then apply bayesian statistic to infer the model parameters favoured by the 
current experimental data, in particular the DAMA annual modulation and Xenon100 exclusion limit. The latter 
strongly disfavours asymmetric scalar doublet DM of mass $\mathcal{O}(\TeV)$ as required by DM-$\overline{\rm DM}$ oscillations, while an asymmetric vector like fermion 
doublet DM with mass around 100 GeV is a good candidate for DAMA annual modulation yet satisfying the 
constraints from Xenon100 data.
\end{abstract}

\maketitle

\section{Introduction}

\begin{figure}[t]
\centering
\includegraphics[width=0.5\columnwidth]{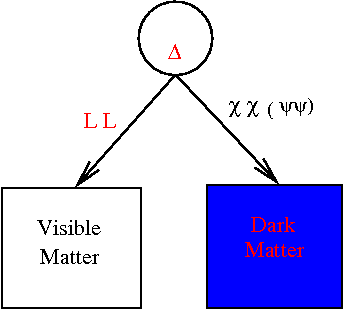}
\caption{A pictorial depiction of triplet scalar $\Delta$ partial decay giving rise to common origin of asymmetries in the lepton and DM sectors.
\label{fig:asymm}}
\end{figure}

The existence of dark matter (DM) is supported by strong gravitational evidences, {\it i.e.} from 
galaxy rotation curves, lensing and large scale structures. This implies that the DM particle
should be electrically neutral, massive and stable on cosmological time scales. However its intrinsic
properties are largely unknown and lead to physics beyond the Standard Model (SM), based on the 
gauge group $SU(3)_C\times SU(2)_{L}\times U(1)_Y$. On the contrary, its relic abundance,  $\Omega_{\rm DM} \sim 0.23$, is well 
measured by the WMAP satellite~\cite{Komatsu:2010fb}. However, the 
mechanism that provides its relic abundance is not yet established. Another issue concerning SM is 
the origin of tiny amount of visible matter in the Universe which is in the form of baryons with 
$\Omega_{\rm b} \sim 0.04$, that could be arising from a baryon asymmetry $n_B/n_\gamma \sim 6.15 \times 10^{-10}$, 
as established by WMAP combined with the big-bang nucleosynthesis (BBN) measurements.

In the standard cosmological picture the early Universe has gone through a period of inflation and then 
reheated to a temperature at least larger than the epoch of BBN. Therefore the observed baryon asymmetry 
of the Universe (BAU) and the DM component must have been generated in the thermal bath after reheating. 
If the reheating temperature is less than electroweak (EW) scale ($\mathcal{O}(100) \GeV$), then it is difficult to generate both DM 
and BAU~\cite{Kohri:2009ka}. On the other hand, if the reheating temperature is larger 
than the EW scale, then a handful of mechanisms are available which can give rise to required BAU, 
while leaving a large temperature window for creating $\Omega_{\rm DM}$ from the DM species, which is 
set by freeze-out, a rather independent mechanism with respect to the dynamics of generating BAU. Indeed 
in most of the cosmological model, the energy density of baryons and of DM are independently determined. 

The fact that the energy density of DM is about a factor of 5 with respect to the baryonic one could be a hint
that both sectors share a common origin and the present relic density of WIMP is also generated by an
asymmetry. Over the years a large number of possibilities for asymmetric DM have been
proposed~\cite{Nussinov:1985xr,Barr:1990ca,Dodelson:1991iv,Kaplan:1991ah,Kuzmin:1996he,Fujii:2002aj,Oaknin:2003uv,Hooper:2004dc,Kitano:2004sv,Cosme:2005sb,Farrar:2005zd,Roszkowski:2006kw,Kaplan:2009ag,Kohri:2009yn,An:2009vq,Frandsen:2010yj,Feldstein:2010xe,An:2010kc,Cohen:2010kn,Shelton:2010ta,Davoudiasl:2010am,Gu:2010ft,Blennow:2010qp,McDonald:2010rn,Hall:2010jx,Dutta:2010va,Haba:2011uz,Falkowski:2011xh,Chun:2011cc,Kang:2011wb,Heckman:2011sw,Frandsen:2011kt,Buckley:2011kk,Iminniyaz:2011yp,MarchRussell:2011fi,DelNobile:2011je}. The most stringent constraint on asymmetric DM candidate comes from
neutron stars and white dwarfs in globular cluster, which exclude asymmetric scalar DM below 
$16$ GeV~\cite{McDermott:2011jp,Kouvaris:2011fi}.

In this paper we consider a relatively heavy asymmetric scalar doublet DM (SDDM) whose stability is provided by 
a remnant $Z_2$ flavour symmetry inspired by the Inert Doublet Model~\cite{Deshpande:1977rw,Ma:2006km,Barbieri:2006dq,LopezHonorez:2006gr,Hambye:2007vf}. Indeed the asymmetry in this model is rather natural: in the limit in which the number violating coupling of DM to Higgs goes to zero the $Z_2$ symmetry that protect the DM is elevated to a global $U(1)_{\rm PQ}$ Peccei-Quinn symmetry. 
We show that the observed relic abundance of SDDM and BAU originate naturally in a type II seesaw 
scenario~\cite{Magg:1980ut,Cheng:1980qt,Schechter:1980gr,Gelmini:1980re,Lazarides:1980nt,Mohapatra:1980yp}, as pictorially depicted in figure~\ref{fig:asymm}. To accomplish the 
unification of asymmetries we extended the SM to include a $SU(2)_L$ scalar triplet $\Delta$ and an inert 
scalar doublet $\chi$. The partial decay width: $\Delta \to LL$, where $L$ is the $SU(2)_L$ lepton doublet, and $\Delta 
\to \chi\chi$ then induce the asymmetry simultaneously in both sectors. The lepton asymmetry is then 
transferred to a baryon asymmetry through the sphaleron transitions. In the low 
energy effective theory the induced vacuum expectation value (vev) of the same Higgs triplet gives rise 
to sub-eV Majorana masses, as required by oscillations experiments~\cite{Fukuda:2001nk,Ahmad:2002ka,Eguchi:2003gg}, 
to the three active neutrinos through the lepton number violating interaction $\Delta L L + 
\Delta^\dagger H H$, where $H$ is the SM Higgs. Thus a triple unification of neutrino mass, 
asymmetric DM and baryon asymmetry of the Universe is achieved in a type-II seesaw scenario.

We show that in case of a SDDM $\chi_0$, the neutral component of the inert scalar doublet $\chi$,
the asymmetry in the DM sector gets washed out below EW phase transition by fast oscillations between $\chi_0$ and its complex conjugate field $\overline{\chi_0}$. This sets
a limit of the mass scale of $\chi_0$ to be $M_{\chi_0} \gwig 2 \TeV$, so that the DM freezes out before oscillations
begin to occur. Such heavy asymmetric SDDM are quite natural to explain positron anomalies at PAMELA and FermiLAT,
while suppressing non-observation of antiproton fluxes~\cite{Nezri:2009jd}. The small number violating quartic coupling $\lambda_5$ of $\chi$ (see section 2.2) to the SM Higgs naturally provides $\mathcal{O}(\rm{keV})$ mass splitting between DM particle and its excited state leading to inelastic 
interaction of detector nuclei and DM. Indeed by definition during 
inelastic scattering a DM particle that scatters off a nucleus produces a heavier state. It has been 
introduced by~\cite{TuckerSmith:2001hy} for reconciling the annual modulation at DAMA~\cite{Bernabei:2010mq} 
experiment and the null results at other experiments, {\it e.g.}~\cite{MarchRussell:2004uf,MarchRussell:2008dy,Arrenberg:2008wy,Cui:2009xq,Alves:2009nf,Finkbeiner:2009mi,Finkbeiner:2009ug,Farina:2011bh,Schwetz:2011xm}. Here we re-investigate~\cite{Arina:2009um} the compatibility of the 
SDDM, explaining the DAMA signal with the most recent exclusion bounds of 
CDMS-II~\cite{Ahmed:2010hw}, CRESST-II~\cite{Angloher:2008jj} and Xenon100~\cite{Aprile:2011ts}. The SDDM appears to be strongly constrained by the new Xenon100 data.
 
In analogy to SDDM, we discuss a 
similar model where the DM candidate is given by a vector like fermionic doublet, odd under a $Z_2$ flavour symmetry, 
with mass ${\cal O}(100) \GeV$, and hence to be called fermion doublet DM (FDDM). It will arise that 
the asymmetric inelastic FDDM is appropriate to explain the high precision annual modulation at DAMA while satisfying the latest 
constraint from Xenon100 experiment. In that case the small mass splitting arises through a small 
Majorana mass of the dark fermion doublet given by the triplet $\Delta$.

The outline of the paper is as follows. The next section presents the particle physics content
of the scalar triplet model which achieve a triple unification of neutrino mass, asymmetric inert 
doublet (scalar and fermion) DM and the observed BAU. After briefly commenting about the 
neutrino mass we describe the most general scalar potential for triplet scalar, SDDM and 
SM Higgs. We also address the issue of generating asymmetries in case of a vector like FDDM model. 
In section~\ref{sec:constraints} we broadly discuss the constraints on asymmetric doublet 
scalar and fermion dark matter. In section~\ref{sec:lepto} the asymmetries in the baryonic and DM 
sector are computed and possible wash-out mechanisms are discussed. The Boltzmann equations are solved numerically using Monte-Carlo-Markov-Chain (MCMC) techniques with CP asymmetries and branching 
fractions as free parameters of the theory. Section~\ref{sec:inel} presents the constraints for 
inelastic scattering on the model parameter space from the current direct search experiments, using 
bayesian inference and marginalising over the velocity distribution of the DM particles. We then come 
to the concluding remarks in section~\ref{sec:concl}. The technical details about bayesian analysis and MCMC are left to~\ref{app1}. In~\ref{app2} we show the results of the analysis of inelastic scattering in the case of the standard maxwellian halo and fixed astrophysical parameters.  

\section{Scalar Triplet Model providing darko-lepto-genesis with Non-zero Neutrino Masses}\label{sec:model}

It is known that the bilinear L-violating coupling ($\Delta L= 2$) of scalar triplet to lepton and Higgs leads to neutrino mass via type II seesaw. Moreover, the out-of-equilibrium decay of triplets through the same coupling can give rise to lepton asymmetry in the early Universe~\cite{Ma:1998dx,Hambye:2000ui}. Here the additional decay of 
scalar triplets to a pair of inert scalar doublets ($\chi$) or a pair of vector like inert 
fermion doublets ($\psi$) simultaneously explain the asymmetries in visible and dark sector. In our convention the scalar triplet is defined as $\Delta = (\Delta^{++}, \Delta^{+},\Delta^0)$, with hypercharge $Y=1$.

\subsection{Triplet Seesaw and Non-zero Neutrino Masses}

Since the lepton number is a conserved quantum number within the SM, the masses of neutrinos are 
exactly zero upto all orders in perturbation theory. On the other hand,
oscillation experiments confirm that the neutrinos are massive, however small, and hence they
mix among themselves. This non-trivial result can be minimally explained by incorporating a heavy 
triplet scalar $\Delta$ to the SM of particle physics. The lepton number violating ($\Delta L= 2$) interaction of $\Delta$ with SM fields is given by the Lagrangian:
\be
\mathcal{L} \supset M_\Delta^2 \Delta^\dagger \Delta + \frac{1}{\sqrt{2}} \left[ \mu_H
\Delta^\dagger H H + f_{\alpha \beta} \Delta L_{\alpha} L_{\beta} + {\rm h.c.} \right]\,,
\label{Lag-1}
\ee
where $H$ and $L$ are the SM Higgs and lepton doublets respectively. After the EW
phase transition $\Delta $ acquires a small induced vev, given by
\begin{equation}
\langle \Delta \rangle = -\mu_H \frac{v^2}{\sqrt{2} M_\Delta^2},
\label{Delta-vev}
\end{equation}
where $v=\langle H \rangle = 246$ GeV. The vev of $\Delta$ is required to satisfy
\be
\rho \equiv \frac{M_W^2}{M_Z^2 \cos^2 \theta} = \frac{1 + 2 x^2}{1+4 x^2}\approx 1\,,
\ee
where $x=\langle \Delta \rangle/v$. The above constraint implies that $\langle \Delta \rangle < {\cal O} (1) \GeV$. The trilinear coupling $\Delta L L$ then give rise to Majorana mass matrix for three 
flavours of light neutrinos as:
\begin{equation}
\left( M_\nu \right)_{\alpha \beta} = \sqrt{2} f_{\alpha \beta} \langle \Delta \rangle
=f_{\alpha \beta} \left( \frac{-\mu_H v^2}{M_\Delta^2} \right) \,.
\end{equation}
Hence for $\langle \Delta \rangle < {\cal O} (1) \GeV$ a wide range of allowed values of $f_{\alpha \beta}$ 
gives rise to required neutrino masses. For $f_{\alpha \beta} \approx {\cal O}(1)$, the required value of 
$\langle \Delta \rangle$ satisfying neutrino masses can be obtained by choosing $\mu_H \sim M_\Delta \sim 10^{12}$ GeV. This implies that the scale of lepton number violation is very high. However, in presence 
of an extra scalar triplet the lepton number violating scale can be brought down to TeV scales without finetuning, so that its dilepton signatures can be studied at LHC~\cite{Sahu:2007uh,McDonald:2007ka,Majee:2010ar}.

\subsection{Inelastic SDDM in Triplet Seesaw Model}\label{sec:SDDM}

We now extend the Lagrangian (\ref{Lag-1}) by including a Inert scalar doublet $\chi \equiv
(\chi^+ \chi^0)^T$ and impose a $Z_2$ symmetry under which $\chi$ is odd while all other
fields are even. As a result $\chi$ does not couple to SM fermions and hence serve
as a candidate of DM. The interactions between $\Delta$, $\chi$ and $H$ can be given by the
scalar potential:
\begin{eqnarray}
V(\Delta, H, \chi) &=&  M_\Delta^2 \Delta^\dagger \Delta + \lambda_\Delta (\Delta^\dagger \Delta)^2
+ M_H^2 H^\dagger H + \lambda_H (H^\dagger H)^2 \nonumber\\
&+& M_\chi^2 \chi^\dagger \chi + \lambda_\chi (\chi^\dagger\chi)^2
+ \left[ \mu_H \Delta^\dagger H H + \mu_\chi \Delta^\dagger \chi \chi + {\rm h.c.}\right]\nonumber\\
&+& \lambda_3 |H|^2 |\chi|^2 + \lambda_4 |H^\dagger \chi|^2
+ \frac{\lambda_5}{2} \left[ (H^\dagger \chi)^2 + {\rm h.c.} \right]\,,
\label{scalar-potential}
\end{eqnarray}
where we have neglected the quartic terms involving $\Delta-H-\chi$ as those are
not relevant for our discussion since the vev of $\Delta$ is small. The vacuum stability of the
potential requires $\lambda_\Delta, \lambda_H, \lambda_\chi > 0$ and $\lambda_L \equiv
\lambda_3 + \lambda_4 -|\lambda_5| > - 2\sqrt{\lambda_\chi \lambda_H}$. We further assume that
$M_\chi^2 > 0$, so that $\chi$ does not develop any vev. This is required in order to distinguish
the visible matter from DM given by the neutral component of the doublet $\chi$. Hence
the true vacuum of the potential is given by:
\be
\langle H \rangle = v;~~~  \langle \chi \rangle = 0~~~ {\rm and}\ \  \ \langle \Delta \rangle = u\,.
\ee
Since $\Delta$ is heavy, its vev is small as demonstrated by Eq.~\ref{Delta-vev} and hence
does not play any role in the low energy dynamics. Therefore, in what follows we neglect
the dynamics of $\Delta $ in low energy phenomena. The perturbative expansion of the fields around the minimum is:
\be
H = \pmatrix{0\cr\\ \frac{v + h }{\sqrt{2}}} ~~~{\rm and}~~~ \chi= \pmatrix{\chi^+\cr\\ \frac{S + i A}
{\sqrt{2}}}\,.
\ee
Thus the low energy spectrum of the theory constitutes two charged scalars $\chi^\pm$, two real
scalars $h,S$ and a pseudo scalar $A$, whose masses are given by:
\begin{eqnarray}
M_{\chi^\pm}^2 &=& M_\chi^2 + \lambda_3 \frac{v^2}{2}\,,\nonumber\\
M_h^2  &=& 2 \lambda_H v^2 \,,\nonumber\\
M_S^2 &=& M_\chi^2 + (\lambda_3 + \lambda_4 + \lambda_5) \frac{v^2}{2}\,,\nonumber\\
M_A^2 &=&  M_\chi^2 + (\lambda_3 + \lambda_4 - \lambda_5) \frac{v^2}{2}\,.
\end{eqnarray}
Depending on the sign of $\lambda_5$, either $S$ or $A$ constitutes the DM. Let us assume that
$\lambda_5$ is negative and hence $S$ is the lightest scalar particle. The next to lightest
scalar particle is then $A$. The mass splitting between them is given by
\be
\Delta M^2 \equiv M_S^2 -M_A^2 = \lambda_5 v^2\,.
\ee
From which we can deduce the coupling
\be
\lambda_5 = \frac{2 M_S \delta}{v^2}\,,
\ee
where $\delta= M_S - M_A$. This plays a key role in the direct searches of DM as we will discuss
later. Notice that in the limit $\lambda_5 \to 0$ in the scalar potential (\ref{scalar-potential}),
there is no mass splitting between $S$ and $A$ and the two degrees of freedom can be re-expressed as
$\chi_0$ and its complex conjugate $\bar{\chi_0}$. In this limit we discuss the asymmetry between
$\chi_0$ and $\bar{\chi_0}$ via the decay of the triplet $\Delta$. We then derive upper bound on
DM number violating processes, namely $\chi \chi \to H^\dagger H^\dagger $ involving the
coupling $\lambda_5$. The smallness of $\lambda_5$ can be attributed to the breaking of a global
$U(1)_L$ symmetry under which $\chi$ carries a lepton number +1. Indeed in the absence of $\mu_H \Delta^\dagger H H$ and $\lambda_5 (H^\dagger \chi)^2+ {\rm h.c.}$ terms in
the Lagrangian, there is a global $U(1)_L$ symmetry. The other parameters $\mu_H$ and $\mu_\chi$,
which involves in the DM number violating processes $\chi \chi \to \Delta \to H H$ and
$\chi \chi \to H \to \bar{f} f$, $f$ being the SM fermion, are not necessarily to be small
as these processes are suppressed by the large mass scale of $\Delta$.

\subsection{Inelastic FDDM in Triplet Seesaw Model}\label{sec:FDDM}

Let us replace the inert scalar doublet $\chi $ by a vector like fermion doublet $\psi \equiv 
(\psi_{\rm DM}, \psi_-)$ of hypercharge $Y=-1/2$. With the same $Z_2$ symmetry, under which $\psi$ is 
odd, the neutral component of $\psi$ {\it i.e.} $\psi_{\rm DM}$ can be a candidate of DM. The relevant 
Lagrangian including the triplet scalar $\Delta$ is:
\be
\fl -\mathcal{L} \supset M_\Delta^2 \Delta^\dagger \Delta + M_D \overline{\psi} \psi + \frac{1}{\sqrt{2}} \left[ \mu_H
\Delta^\dagger H H + f_{\alpha \beta} \Delta L_{\alpha} L_{\beta} + g \Delta \psi \psi
+ {\rm h.c.} \right]\,,
\label{Lag-DM}
\ee
where $M_D \sim {\cal O}(100) \GeV $ is the Dirac mass of $\psi$. The bilinear DM coupling $\Delta \psi\psi$ can be re-expressed as:
\begin{eqnarray}
\fl \frac{1}{\sqrt{2}} g\Delta \psi \psi  & \equiv &  \frac{1}{\sqrt{2}} g \overline{\psi^c} i \tau_2 \Delta \psi \nonumber\\
 &  = &   -\frac{1}{2} g \left[ \sqrt{2} (\overline{\psi_-^c} \psi_- \Delta^{++}) + (\overline{\psi_-^c} \psi_{\rm DM} + 
\overline{\psi_{\rm DM}^c}\psi_- )\Delta^+
 - \sqrt{2} ( \overline{ \psi_{\rm DM}^c} \psi_{\rm DM} \Delta^0) \right]\,, \nonumber \\ 
\end{eqnarray}
where we have used the matrix form of the triplet scalar: 
\be
\Delta = \pmatrix{ \frac{\Delta^+}{\sqrt{2}} & \Delta^{++}\cr
\Delta^0 &  -\frac{\Delta^+}{\sqrt{2}} }\,.
\ee
After EW symmetry breaking the neutral component of $\Delta$ acquires an induced vev and hence give rise 
a small Majorana mass to $\psi$, $ m = \sqrt{2} g \langle \Delta^0 \rangle$.

Therefore the Dirac spinor $\psi_{\rm DM}$ can be written as sum of two Majorana spinors $(\psi_{\rm DM})_L$ and 
$(\psi_{\rm DM})_R$. The Lagrangian for the DM mass becomes: 
\bea
-\mathcal{L}_{\rm DM mass} &=& M_D \left[ \overline{(\psi_{\rm DM})_L} (\psi_{\rm DM})_R 
+ \overline{ (\psi_{\rm DM})_R} (\psi_{\rm DM})_L \right] \nonumber\\
&& + m \left[ \overline{ (\psi_{\rm DM})_L^c} (\psi_{\rm DM})_L + \overline{ (\psi_{\rm DM})_R^c} (\psi_{\rm DM})_R \right] \,.
\eea
This implies there is a $2\times 2$ mass matrix for the DM in the basis $\{(\psi_{\rm DM})_L, 
(\psi_{\rm DM})_R\}$. By diagonalising it two mass eigenstates 
$(\psi_{\rm DM})_1$ and $(\psi_{\rm DM})_2$ arise, with masses $M_{\psi_1}=M_D -m$ and $M_{\psi_2}=M_D + m$. Thus the mass 
difference between the two states $\delta = 2 m \sim {\cal O} (100) \keV$ is required by the direct 
search experiments. We will come back to this issue while discussing inelastic scattering of DM with nucleons. Note that in this case the 
inelastic scattering of DM with nucleons ( {\it i.e.} $(\psi_{\rm DM})_1 N \to (\psi_{\rm DM})_2 N $) via SM $Z$-exchange 
dominates to elastic scattering, as in the case of the scalar candidate. 

Now we will briefly comment about the dark matter asymmetry. Similar to the decay of 
$\Delta \to \chi \chi $, 
the decay of $\Delta \to \psi \psi $ will produce an asymmetry in the dark sector. Since $\psi$ is 
odd under a $Z_2$ flavor symmetry, it will not couple to any other SM fields and hence 
the asymmetry will remain in $\psi_{\rm DM}$ for ever, namely in this case there are no strong wash-out processes.

\subsection{Sub-eV Neutrino Mass versus keV Majorana mass of FDDM}

Notice that in case of FDDM, the induced vev of $\Delta$ introduces two mass scales. One is the Majorana mass 
of neutrinos, {\it i.e.} $M_\nu = \sqrt{2} f \langle \Delta^0 \rangle \sim  {\cal O}(1) \eV$ and other is the 
Majorana mass of DM, {\it i.e.} $m = \sqrt{2} g \langle \Delta^0 \rangle \sim {\cal O}(100) \keV$. This implies 
a hierarchy between the two couplings $f$ (third term in Eq.~\ref{Lag-DM}) and $g$ (fourth term in Eq.~\ref{Lag-DM}) 
of the order of ${\cal O}(10^5)$ in order to explain the triple unification of neutrino mass, asymmetric DM 
and BAU. 

\section{Constraints on Asymmetric Inert Doublet (Scalar and Fermion) DM}\label{sec:constraints}

\subsection{Constraints on SDDM from Oscillation}
In case of SDDM the two states $\chi_0$ and its complex conjugate $\bar{\chi}_0$ 
can be written in terms of the mass eigenstates $S$ and $A$:
\bea
| \chi_0 \rangle  &=& \frac{1}{\sqrt{2}} (S + i A)\,, \nonumber\\
| \bar{\chi}_0 \rangle &=& \frac{1}{\sqrt{2}} (S - i A)\,.
\label{flavor-states}
\eea
The state $| \chi_0 \rangle $ at any space-time point $(x,t)$ is given by
\be
|\phi(x,t) \rangle  = \frac{1}{\sqrt{2}} \left[ e^{-i(E_S t -k_S x)} | S \rangle  + i e^{+i(E_A t- k_A x)}
| A \rangle \right]\,,
\label{wavefunction}
\ee
where $E_S=\sqrt{k_S^2 + M_S^2 }$ and $E_A=\sqrt{k_A^2 + M_A^2 }$ are the energy of $S$ and $A$
respectively. The probability of $| \chi_0 \rangle $ oscillating into $| \bar{\chi}_0 \rangle$
is then given by 
\be
P_{| \chi_0 \rangle  \to | \bar{\chi}_0 \rangle } = |\langle \bar{\chi}_0| \phi(x,t) \rangle |^2 \,.
\ee
Using Eqs.~\ref{flavor-states} and~\ref{wavefunction} the probability of oscillation takes the form:
\be
P_{|\chi_0 \rangle  \to | \bar{\chi}_0 \rangle } = \frac{1}{4} \left[ 2 - e^{-i\left[(E_S-E_A)t
- (k_A-k_S)x \right]} - e^{+i\left[(E_S-E_A)t - (k_A-k_S)x \right]} \right]\,.
\label{probability}
\ee

Above the EW phase transition there is no mass splitting between the two mass eigenstates $S$ and $A$, therefore $M_S=M_A$, $E_S=E_A$ and $k_S=k_A$. As a result from
Eq.~\ref{probability} the probability of oscillation is null:
\be
P_{|\chi_0 \rangle  \to | \bar{\chi}_0 \rangle } = 0\,.
\ee

Below the EW phase transition the DM number violating term $\frac{\lambda_5}{2}\left( (H^\dagger \chi)^2
+ {\rm h.c.} \right)$ produce a mass splitting between the two mass eigenstates $S$ and $A$. From Eq.~\ref{probability} the
probability of oscillation becomes:
\be
P_{|\chi_0 \rangle  \to | \bar{\chi}_0 \rangle } \simeq \frac{1}{2} \left[ 1- \cos
\left(\frac{\Delta M^2 (t-t_{\rm EW})}{2 E} \right) \right]\,,
\label{probability_EW}
\ee
where we have assumed $E_S \sim E_A \sim E$, which is a good approximation for a small mass
splitting. In the following we will consider a mass splitting of ${\cal O}({\rm keV})$,
which implies $\lambda_5 \sim 10^{-7}$. We also normalise the time of evolution from the time of
EW phase transition, so that at $t=t_{\rm EW}$, $P_{|\chi_0 \rangle  \to | \bar{\chi}_0 \rangle }=0$.
Below EW phase transition the time of oscillation from $\chi_0$ to $\bar{\chi}_0$ can be
estimated as
\be
t-t_{\rm EW} = \frac{2 E \pi}{\Delta M^2}\,.
\ee
In the relativistic limit the energy of the DM particle $E \sim T $, where $T$ is the temperature of
the thermal bath. Hence the oscillation time can be given as:
\be
t-t_{\rm EW} \sim  4 \times 10^{-10} {\rm s} \left( \frac{T}{100 {\rm GeV}} \right) \left( \frac{{\rm keV}^2}
{\Delta M^2} \right) \,.
\ee
On the other hand, in the non-relativistic limit the energy of the DM particle $E \sim M_S$. Thus
for $M_S \sim 100\  {\rm GeV}$, the time of oscillation is again similar to relativistic case.
This implies that $\chi_0$ oscillates rapidly to $\bar{\chi}_0$. In this case if $\chi_0$ is in
thermal equilibrium then during each oscillation there is a leakage of asymmetry through the
annihilation channel $\chi_0 \bar{\chi}_0 \to {\rm SM\  particles}$. Alternatively to keep the
generated asymmetry intact $\chi_0$ should freeze-out before it oscillate to $\bar{\chi}_0$.
In other words, the mass of $\chi_0$ should be given by
\be
M_{\chi_0} \gae x_f T_{\rm EW}\,,
\ee
where $x_f \sim 20 $, which determines the epoch of freeze-out. From the above equation
we see that to get an asymmetric SDDM one should have $M_{\chi_0} \gae 2\  {\rm TeV}$.

\subsection{Constraints from Collider}

Since the DM (scalar or fermion) is a doublet under the SM gauge group, it couples to 
the $Z$ boson. As a result they can change the invisible decay width of the later unless 
the mass of DM is greater than half of $Z$-boson mass. This gives a lower bound on the 
mass scale of either SDDM or FDDM to be $\gwig 45 \GeV$~\cite{Lundstrom:2008ai}.

\section{Developing asymmetries in the lepton and DM sectors}\label{sec:lepto}

\begin{figure}[t]
\centering
\includegraphics[width=0.5\columnwidth]{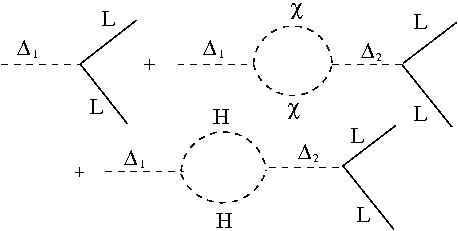}
\caption{Tree level and self energy correction diagrams for the production of CP asymmetry in
leptogenesis.
\label{fig-2}}
\end{figure}

\begin{figure}[t]
\centering
\includegraphics[width=0.5\columnwidth]{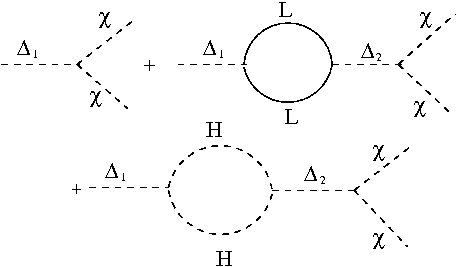}
\caption{Tree level and self energy correction diagrams for the production of CP asymmetry in
generating asymmetric DM.
\label{fig-3}}
\end{figure}

If the triplet $\Delta$ is heavy enough as required by the seesaw, then it can go
out-of-equilibrium even if the gauge couplings are ${\cal O}(1)$~\cite{Ma:1998dx,Hambye:2000ui,Hambye:2003ka,Sahu:2006pf}. In such a case the out-of-equilibrium decays of $\Delta \to LL$
and $\Delta \to \chi\chi \ (\psi \psi)$ produce asymmetries in visible and dark sectors respectively.
The CP asymmetry for the two sectors arise via the interference of tree level decay and
self-energy correction diagrams as shown in figure~\ref{fig-2} and~\ref{fig-3}
respectively, for the scalar DM case, but totally analogous for the fermionic doublet.

Considering the inert scalar doublet as a reference for the scalar potential, from Figs.~\ref{fig-2} and~\ref{fig-3} we see that the CP asymmetry requires at least two
triplet scalars. Hence in presence of their interactions, the diagonal mass $M_\Delta^2$ in
Eq.~\ref{scalar-potential} is replaced by:
\be
\frac{1}{2} \Delta_a^\dagger \left( {\mathcal M}_+^2 \right)_{ab} \Delta_b + \frac{1}{2}
(\Delta^*_a)^\dagger \left( {\mathcal M}_-^2 \right)_{ab} \Delta_b^*\,,
\label{mass-matrix}
\ee
and the trilinear couplings $\mu_H \Delta^\dagger H H  + \mu_\chi \Delta^\dagger \chi \chi
+ {\rm h.c.}$ in the scalar potential (\ref{scalar-potential}) become:
\be
\sum_{a=1,2}\mu_{aH} \Delta^\dagger H H  + \mu_{a\chi} \Delta^\dagger \chi \chi + {\rm h.c.}\,.
\ee
In Eq.~\ref{mass-matrix}, the mass matrix is given by:
\begin{equation}
{\mathcal M}_\pm^2 = \pmatrix{M_1^2-i C_{11} & -i C_{12}^{\pm} \cr\\
-i C_{21}^\pm & M_2^2- i C_{22} }\,,
\label{mass-matrix-1}
\end{equation}
where
\be
C_{ab}^+ = \Gamma_{ab} M_b = \frac{1}{8\pi}\left(\mu_{aH}\mu_{bH}^* + \mu_{a\chi}\mu_{b\chi}^*+
M_a M_b \sum_{\alpha\beta}f_{a\alpha\beta}^* f_{b\alpha\beta} \right)\,,
\ee
with $C_{ab}^- = \Gamma_{ab}^* M_b$, and $C_{aa} = \Gamma_{aa} M_a$. Solving the mass matrix~\ref{mass-matrix-1} one gets two mass eigenstates $\xi^+_{1,2}=A_{1,2}^+ \Delta_1 + B_{1,2}^+
\Delta_2$ with masses $M_1$ and $M_2$. The complex conjugate of $\xi^+_{1,2}$ are given by 
$\xi^-_{1,2}=A_{1,2}^- \Delta_1 + B_{1,2}^- \Delta_2$. Note that $\xi^+$ and $\xi^-$ states are 
not CP eigenstates and hence their decay can give rise to CP asymmetry. We assume that there is 
no asymmetry, either in the visible sector or in the dark sector, at a temperature above the mass 
scale of the triplets. The asymmetries are generated in a thermal bath by the decay of these triplets. 
If we further assume that the mass of $\xi^\pm_{1}$ is much less than the mass of $\xi^\pm_{2}$ then the 
final asymmetries in visible and dark sectors will be given by the decay of $\xi^{\pm}_{1}$ as:
\begin{eqnarray}
\epsilon_L &=& 2 \left[ {\rm Br}(\xi_1^{-}\to \ell \ell) - {\rm Br} (\xi_1^{+} \to
\ell^c \ell^c) \right] \equiv \epsilon_{\rm vis}\,,  \nonumber\\
\epsilon_{\chi} &=& 2 \left[ {\rm Br}(\xi_1^{-}\to \chi_0 \chi_0) - {\rm Br} (\xi_1^{+}
\to \chi_0^* \chi_0^*) \right] \equiv \epsilon_{\rm dark} \,,
\end{eqnarray}
where the front factor 2 takes into account of two similar particles are produced per decay.
From Figs.~\ref{fig-2} and~\ref{fig-3}, the asymmetries are estimated to be:
\be
\epsilon_{L} = \frac{{\rm Im} \left( \mu_{1\chi} \mu_{2\chi}^* \left[1 + \frac{\mu_{1H} \mu_{2H}^*}
{\mu_{1\chi} \mu_{2\chi}^* } \right] \sum_{\alpha \beta} f_{1\alpha\beta} f^*_{2\alpha\beta}
\right) } {8\pi^2 (M_2^2- M_1^2)} \left[\frac{M_1}{\Gamma_{1}} \right]\,,
\label{cp_vis}
\ee
and
\be
\epsilon_{\chi} = \frac{{\rm Im} \left( \mu_{1\chi} \mu_{2\chi}^* \left[ \frac{\mu_{1H} \mu_{2H}^*}
{M_1^2} + \sum_{\alpha \beta} f_{1\alpha\beta} f^*_{2\alpha\beta}
\right] \right) } {8\pi^2 (M_{2}^2- M_{1}^2)} \left[\frac{M_1}{\Gamma_{1}} \right]\,,
\label{cp_dark}
\ee
where $\Gamma_1 \equiv \Gamma_{11}$.

In a thermal bath these asymmetries evolve as the Universe expands and settle to a
final value as soon the relevant processes go out of equilibrium, {\it i.e.}
\begin{equation}
\Gamma_i \equiv n_i \langle \sigma_i|v| \rangle \ll  H(T)\,,
\end{equation}
where $H(T)$ is the Hubble scale of expansion. As a result the Yields in both sectors can
be written as:
\begin{eqnarray}
Y_L \equiv \frac{n_L}{s} = \epsilon_L X_\xi \eta_L\,, \nonumber\\
Y_{\chi} \equiv \frac{n_\chi}{s} = \epsilon_\chi X_\xi \eta_\chi\,,
\label{asymmetry_L_DM}
\end{eqnarray}
where $X_\xi = n_{\xi_1^-}/s \equiv n_{\xi_1^+}/s$, $s=2(\pi^2/45) g_* T^3$ is the entropy
density and $\eta_{L},\eta_{\chi}$ are the efficiency factors, which take into account the 
depletion of asymmetries due to the number violating processes involving $\chi$, $L$ and $H$ (this holds also for the fermionic inert doublet, hence we can replace $\chi \to {\rm DM}$ label). At a 
temperature above the EW phase transition a part of the lepton asymmetry gets converted to the baryon 
asymmetry via the $SU(2)_L$ sphaleron processes. As a result the baryon asymmetry~\cite{Harvey:1990qw} is:
\be
Y_B = -\frac{8 n + 4 m}{14 n + 9 m} Y_L = - \mathcal{S}_{\rm DM} Y_L\,,
\label{B-asy}
\ee
where $n$ is the number of generation and $m$ is the number of scalar doublets, leading to $\mathcal{S}_{\rm DM}=0.53,0.55$ for scalar DM and fermionic DM respectively.

As introduced in Sec~\ref{sec:SDDM}, in the case of the scalar doublet DM, the asymmetry may strongly washed out, if kinematically allowed, by the DM number violating processes:
$\chi \chi \to \Delta \to H H$, $\chi\chi \to H^\dagger H^\dagger$ (contact annihilation through
$\lambda_5$ coupling) and $\chi \chi \to H \to \bar{f} f$. The reduced cross-section for the
former process is given by:
\be
\hat{\sigma} (\chi \chi \to \Delta \to H H) = \frac{1}{8 \pi} \frac{|\mu_\chi|^2 |\mu_H|^2}
{(\hat{s}-M_1^2)^2}\,,
\ee
where $\hat{s}$ is the centre of mass energy for the process: $\chi\chi \to \Delta \to HH$. Below the
mass scale of the triplet this process is strongly suppressed. On the other hand in case of the contact 
annihilation of $\chi$'s the reduced cross-section is given by
\be
\hat{\sigma}_\chi = \frac{\lambda_5^2}{32 \pi}\,.
\label{reduced-cross-section}
\ee
As a result the reaction rate is given by $\Gamma_\chi = (\gamma_\chi/n_\chi^{\rm eq})$, where the reaction density is
\be
\gamma_\chi = \frac{T}{64 \pi^4} \int_{\hat{s}_{\rm min}}^{\infty} d\hat{s} \sqrt{\hat{s}}
K_1\left(\frac{\sqrt{\hat{s}}}{T} \right)\hat{\sigma}_\chi\,,
\label{scattering-density}
\ee
and the equilibrium number density of $\chi$ is
\be
n_\chi^{\rm eq} = \frac{g_{\rm dof} M_\chi^2 T}{2 \pi^2} K_2\left(\frac{M_\chi}{T} \right)\,,
\label{eq_density}
\ee
where $g_{\rm dof}$ is the internal degrees of freedom and $\hat{s}$ is the usual Mandelstam variable for the center of mass energy. In Eqs.~\ref{scattering-density} and~\ref{eq_density}, $K_1$ and $K_2$ are modified Bessel functions.
\begin{figure}[t]
\centering
\includegraphics[width=0.9\columnwidth]{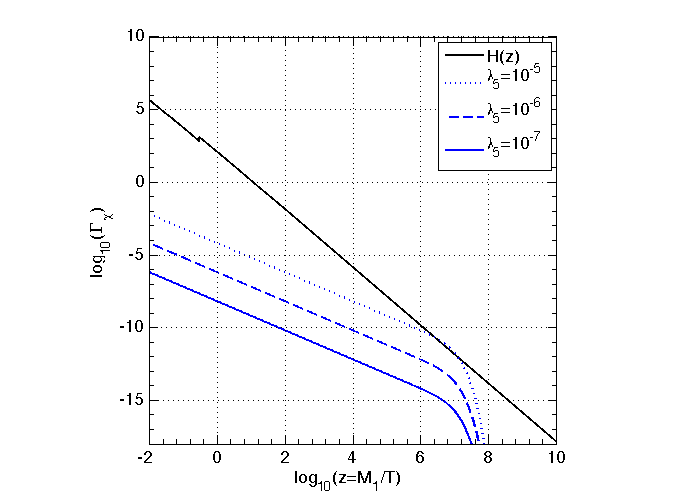}
\caption{The scattering rate of the process $\chi\chi \to H^\dagger H^\dagger$ for different
values of $\lambda_5$ is compared with the Hubble expansion rate. For illustration purpose we have used 
$M_\chi= 2 \TeV$ and $M_1=10^{10}\GeV$.}
\label{rate_Hubble}
\end{figure}
In fig.~\ref{rate_Hubble} we compare the rate of the process $\chi\chi \to H^\dagger H^\dagger$ 
with the Hubble rate by taking three values of $\lambda_5$. We see that for $\lambda_5  \lwig 10^{-5}$ (blue dotted line), the 
scattering rate remains out-of-equilibrium through out the epoch. For larger values of $\lambda_5$ 
the scattering process comes to equilibrium at late epoch. At around $z\equiv M_1/T \approx 10^7$, which implies 
$M_\chi/T = (M_\chi/M_1) z \approx 1$, the scattering rate of the process sharply drops as it is expected 
and does not depend on the value of $\lambda_5$. This argument also holds in case of the process $ \chi \chi \to H \to \bar{f} f$. However, the rate 
of the scattering is further suppressed by the mass scale of Higgs. 

Returning to the general case of doublet DM, both scalar and fermion, from Eqs.~\ref{asymmetry_L_DM} and~\ref{B-asy} the DM to baryon ratio is given by:
\be\label{eq:IMP}
\frac{\Omega_{\rm DM}}{\Omega_B} = \frac{1}{\mathcal{S}_{\rm DM} }\frac{m_{\rm DM}}{m_p} \frac{\epsilon_{\rm DM}}{\epsilon_L}
\frac{\eta_{\rm DM}}{\eta_L}\,,
\ee
where $m_p\sim 1\GeV $ is the proton mass. From this equation it is clear that the dependence of $\eta_{\rm DM}/\eta_L$ on the mass of DM goes as $1/m_{\rm DM}$. Hence for a $\mathcal{O}(100)$ GeV scale 
DM, the required efficiency factor for DM is two orders of magnitude less than the case of lepton provided 
that the CP asymmetries are equal on both sectors (see the end of section~\ref{sec:numsol}). For example, from Eqs.~\ref{cp_vis} and~\ref{cp_dark} 
we notice that the CP asymmetries are identically equal if:
\be
\frac{\mu_{1\chi} \mu_{2\chi}^*}{M_1^2} = \sum_{\alpha,\beta} f_{1\alpha\beta} f^*_{2\alpha\beta}\,.
\ee

In what follows, section~\ref{sec:numsol}, we solve numerically the relevant Boltzmann equations for quasi equilibrium evolution of 
triplet scalars, presented in section~\ref{sec:boleq}, to show that the parameter space of the theory fulfills the criteria $\Omega_{\rm DM} 
\sim 5\  \Omega_B$ and the observed BAU, while allowing a broad range of asymmetric DM masses. 

\subsection{Boltzmann equations for quasi-equilibrium evolution of triplet scalars}\label{sec:boleq}

If the triplet ($\xi_1^\pm$) decay occurs in a quasi-equilibrium state then the detailed of
$\eta_\chi$ and $\eta_L$ depends on the dynamics of the processes occuring in the thermal bath
and can be obtained by solving the relevant Boltzmann equations~\cite{Hambye:2005tk,Chun:2006sp}. In our case the additional decay channel of the scalar triplet into DM particles is included. 

At first the number density of $\xi_1^\pm$ particles changes due to their decay ($\xi_1^\pm \to LL,
HH, \chi\chi $ or $\psi\psi$) and gauge annihilation ($\xi_1^- \xi_1^+ \to \bar{L} L, H^\dagger H, \chi^\dagger \chi (\bar{\psi}\psi),
W^\mu W_\mu, B^\mu B_\mu$), where $W^\mu$ and $B^\mu$ are the $SU(2)_L$ and $U(1)_Y$ gauge bosons
respectively. If we assume that the masses of the components of the triplet are same before EW symmetry breaking then it is fairly general to use the dimensionless variables
$z=M_1/T$ and $X_\xi = n_{\xi_1^-}/s \equiv n_{\xi_1^+}/s$. The Boltzmann equation for the evolution of $\xi_1^\pm$ density is then given by:
\begin{equation}
\frac{dX_{\xi_1}}{dz}=-\frac{\Gamma_D}{zH(z)}\left( X_{\xi_1} -X_{\xi_1}^{\rm eq} \right) -
\frac{\Gamma_A}{z H(z)} \left( \frac{X_{\xi_1}^2-{X_{\xi_1}^{\rm eq}}^2}{X_{\xi_1}^{\rm eq}} \right)\,,
\label{boltzman-1}
\end{equation}
where
\begin{equation}
\Gamma_D=\Gamma_1 \frac{K_1(z)}{K_2(z)}\,,\,\,\Gamma_A = \frac{\gamma_A}{n_{\xi_1}^{\rm eq}}
~~~{\rm and}~~~H(z)= \frac{ H(T=M_1)} {z^2}\,,
\end{equation}
and $\gamma_A$'s are the scattering densities, described in Eqs.~\ref{eq:scats}. The temperature independent decay rate
of $\xi_1$ can be written as a function of the neutrino mass:
\be
\Gamma_1=\frac{1}{8\pi} \frac{|m_\nu| M_1^2}{\langle H \rangle^2 \sqrt{B_L B_H}}\,,
\ee
where $B_L$ and $B_H$ are the branching fractions in the decay channels: $\xi_1 \to L L$ and 
$\xi_1 \to H H$. Note that we have re-expressed the total decay rate $\Gamma_1(f_{\rm DM},f_H,f_L,M_1)$, where $f_{\rm DM} \equiv \mu_\chi/M_1$ for SDDM and $g$ for FDDM and $f_H\equiv \mu_H/M_1$, as $\Gamma_1(m_\nu, B_L, B_H, M_1)$. In the following we set $m_\nu=0.05$ eV and therefore the total decay rate depends only on three variables, namely $B_L$, $B_H$ and $M_1$. This makes a crucial decision in setting up the final asymmetry as we will show in section~\ref{sec:numsol}.

For the gauge annihilation processes, the scattering densities are given by:
\begin{eqnarray}\label{eq:scats}
\fl \gamma ( \xi_1^+\xi_1^-\to \bar{f}f)  &=& \frac{M_1^4 (6 g_2^4 + 5 g_Y^4)}{128 \pi^5 z}
\int_{x_{\rm min}}^{\infty} dx \sqrt{x} K_1(z \sqrt{x}) r^3 \,,\nonumber\\
\fl \gamma ( \xi_1^+\xi_1^-\to H^\dagger H) &=& \frac{M_1^4 (g_2^4 + g_Y^4/2)}{512\pi^5 z}
\int_{x_{\rm min}}^{\infty} dx \sqrt{x} K_1(z \sqrt{x})r^3 \,,\nonumber\\
\fl \gamma ( \xi_1^+\xi_1^-\to \chi^\dagger \chi) &=& \frac{M_1^4 (g_2^4 + g_Y^4/2)}{512\pi^5 z}
\int_{x_{\rm min}}^{\infty} dx \sqrt{x} K_1(z \sqrt{x})r^3 \,,\nonumber\\
\fl \gamma ( \xi_1^+\xi_1^-\to W^a W^b) &=& \frac{ M_1^4 g_2^4}{64 \pi^5 z}
\int_{x_{\rm min}}^{\infty} dx \sqrt{x} K_1(z \sqrt{x})
\left[r (5+34/x)-\frac{24}{x^2}(x-1)\ln \left(\frac{1+r}{1-r} \right)\right]\,,\nonumber\\
\fl \gamma ( \xi_1^+\xi_1^-\to BB) &=& \frac{3 M_1^4 g_Y^4}{128 \pi^5 z}
\int_{x_{\rm min}}^{\infty} dx \sqrt{x} K_1(z \sqrt{x})\nonumber\\
&& \times \left[r(1+4/x)-\frac{4}{x^2}(x-2) \ln \left( \frac{1+r}{1-r} \right) \right]\,,
\end{eqnarray}
where $r= \sqrt{1-4/x}$ and $x=\hat{s}/M_1^2$. In case of FDDM, the process corresponding to $\xi_1^+\xi_1^-\to \chi^\dagger \chi$ is given by:
\begin{equation}
\gamma ( \xi_1^+\xi_1^-\to \bar{\psi}\psi)  = \frac{M_1^4 (6 g_2^4 + 5 g_Y^4)}{128 \pi^5 z}
\int_{x_{\rm min}}^{\infty} dx \sqrt{x} K_1(z \sqrt{x}) r^3 \,.
\end{equation}

Since $\xi_1^\pm$ are charged particles there is an evolution of the asymmetry: $Y_{\xi_1} = 
( n_{\xi_1^-}-n_{\xi_1^+})/s$ due to the decay and inverse decay of $\xi_1^\pm$ particles. The evolution of 
$Y_{\xi_1}$ is described by the Boltzmann equation:
\begin{equation}
\frac{d Y_{\xi_1}}{dz} = -\frac{\Gamma_D}{zH(z)} Y_{\xi_1} + \sum_j \frac{\Gamma^j_{ID}}{zH(z)}
2 B_j Y_j\,,
\label{boltzman-2}
\end{equation}
where $Y_j=(n_j-n_{\bar j})/s$, with $j=L, H, \chi \ (\psi)$ and
\be
\Gamma^j_{ID} = \Gamma_D \frac{X_{\xi_1}^{\rm eq}}{X_j^{\rm eq}}~~~{\rm and}~~~ B_j=\frac{\Gamma_j}
{\Gamma_1}\,,
\label{eq:pippo}
\ee
where $X_j=n_j/s$. The evolution of the asymmetries 
$Y_j$ is given by the Boltzmann equation:
\bea
\frac{d Y_j}{dz} &=&\ 2\  \Big\{ \frac{\Gamma_D}{zH(z)} \left[ \epsilon_j (X_{\xi_1} - X_{\xi_1}^{\rm eq}) \right]
+  B_j \left( \frac{\Gamma_D}{zH(z)} Y_{\xi_1}  - \frac{\Gamma^j_{ID}}{zH(z)} 2 Y_j\right)\nonumber\\
&& -\sum_k \frac{\Gamma^k_S}{z H(z)} \frac{X_{\xi_1}^{\rm eq}}{X_k^{\rm eq}} 2 Y_k\Big\}\,.
\label{boltzman-3}
\eea
where $\Gamma_S=\gamma_S/n_{\xi_1^-}^{\rm eq}$ is the scattering rate involving the number
violating processes, such as $\chi \chi \to \xi \to HH$, $LL \to \xi \to HH$. The front factor in Eq.~\ref{boltzman-3} takes into account
of the two similar particles produced in each decay.

Solving the Boltzmann Eqs.~\ref{boltzman-1}, ~\ref{boltzman-2} and~\ref{boltzman-3} we can get the lepton
($Y_L$) and dark matter ($Y_{\rm DM}$) asymmetries. Note that because of the conservation of
hypercharge the Boltzman equations~\ref{boltzman-1}, ~\ref{boltzman-2} and~\ref{boltzman-3}
satisfy the relation: $2 Y_\xi + \sum_j Y_j =0$.  This implies:
\be
Y_\xi =-\frac{1}{2} \sum_j Y_j\,.
\ee

We will follow a phenomenological approach and calculate the ratio of efficiency factors $\eta_{\rm DM}$ and $\eta_L$ (and hence also the individual efficiency) solving the set of coupled equations~\ref{boltzman-1}, ~\ref{boltzman-2} and~\ref{boltzman-3}. As usual the efficiency factor for the species $i=L,H,{\rm DM}$ is defined as:
\begin{equation}
\eta_i = \frac{Y_i}{\epsilon_i \ X_\xi\Big|_{T >> M_1}}\,.
\end{equation}

The free parameters of the model are the CP asymmetries $\epsilon_i$ for all the species, the dark matter mass $m_{\rm DM}$ and the triplet mass $M_1$. However in the remaining of the paper we will focus on heavy triplet, $M_{1} \sim 10^{10}$ GeV, which has been shown to lead to successful leptogenesis~\cite{Hambye:2005tk,Chun:2006sp} for a wide range of CP asymmetries and branching ratios. In addition the following constraints apply:
\begin{equation}\label{eq:const}
\sum_j \epsilon_j =0\,,  \ \ \
\sum_j B_j =1 \ \ {\rm and}\   | \epsilon_j| \le 2 \ B_j\,.
\end{equation}
The first and third conditions ensure that all amplitudes are physical and the total amount of CP violation can not exceed 100\% in each channel, while the second condition simply demands unitarity of the model. The number of free parameters therefore drops to 5, which we choose to be: $\epsilon_L, \epsilon_{\rm DM}, B_L, B_{\rm DM}$ and $m_{\rm DM}$. The numerical procedure and the results are described in the next section.

\subsection{Numerical solutions of the Boltzmann equations}\label{sec:numsol}

In principle for unlimited computational power one could use a griding method to explore the whole parameter space and to localise 
the hypervolume that satisfies Eq.~\ref{eq:IMP}. For the problem under scrutiny we have a 5-dimensional space: MCMC technique is well suited in this case since it numerically scales linearly with the number of dimension instead of exponentially. We follow the approach presented in~\cite{Clesse:2009ur}~\footnote{MCMC technique has been applied for scanning hybrid inflationary parameter space~\cite{Clesse:2009ur} and for investigating the multi-dimensional parameter space of supersymmetric theories, {\it e.g.}~\cite{deAustri:2006pe}.}.

\begin{figure}[t]
\begin{minipage}[t]{0.5\textwidth}
\centering
\includegraphics[width=0.9\columnwidth]{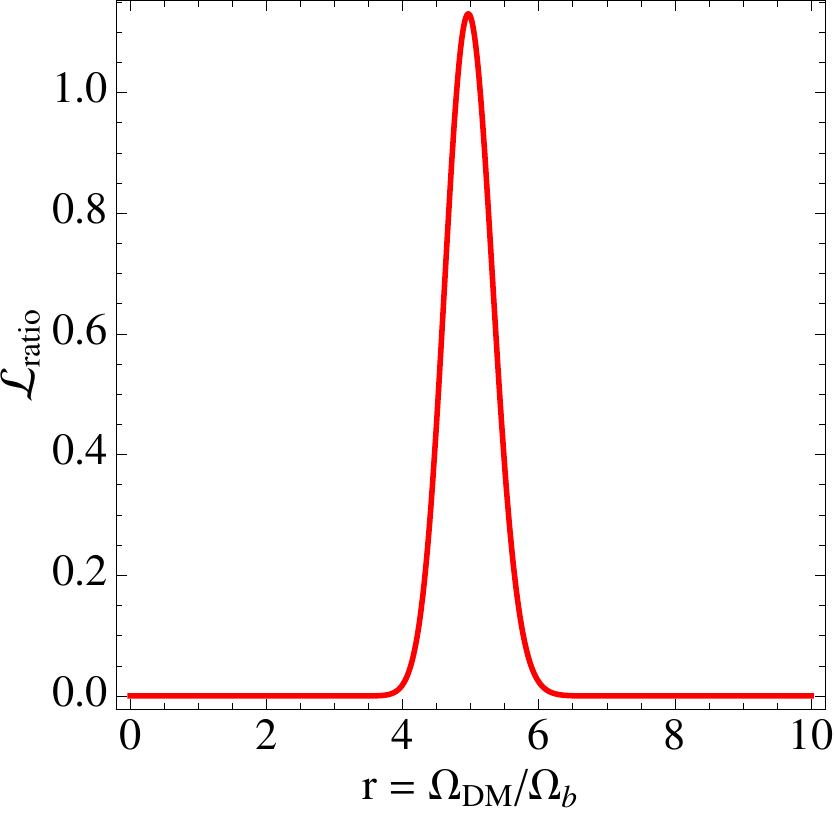}
\end{minipage}
\hspace*{-0.2cm}
\begin{minipage}[t]{0.5\textwidth}
\centering
\includegraphics[width=0.9\columnwidth]{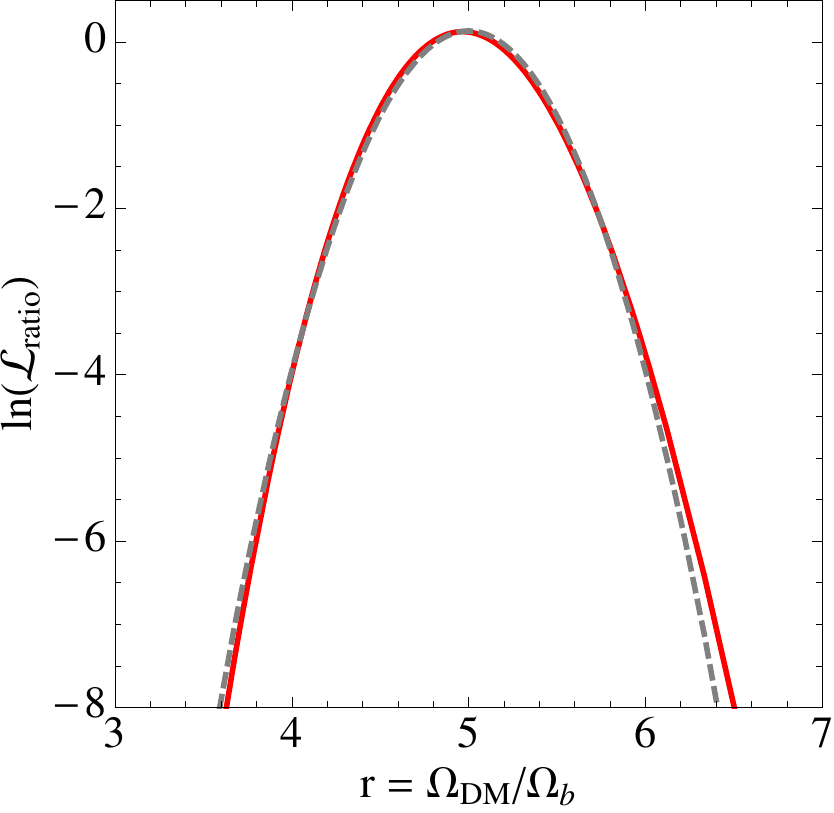}
\end{minipage}
\caption{{\it Left:} Likelihood function $\mathcal{L}_{\rm  ratio}$ given by the ratio distribution, as a function of $r\equiv \Omega_{\rm DM}/\Omega_b$. {\it Right:} Logarithm of the ratio likelihood (red line) and logarithm of a gaussian distribution (dashed gray curve). Both curve have the same variance.}
\label{fig:lkhratio}
\end{figure}

Defining a probability measure over the full parameter space allows to use Bayesian inference to assess the posterior probability distribution of all the parameters to get asymmetry in the dark sector as well 
as in the lepton sector, which gives rise to observed BAU. The details about bayesian statistical methods and the implementation of the MCMC are given in~\ref{app1}, while for detailed reviews we refer to {\it e.g.}~\cite{loredo,Trotta:2008qt}. The important point that has to be underlined is that through Bayes' theorem:
\begin{equation}
\label{eq:bayes}
  \mathcal{P} (\theta | X) {\rm d}\theta \propto\ \mathcal{L}(X |
  \theta) \cdot  \pi(\theta) {\rm d}\theta \,,
\end{equation}
the posterior probability density function (pdf) $\mathcal{P}(\theta|X)$ for the model parameter $\theta$, given the data $X$, is proportional to the likelihood of the experiment times the prior belief in our model $\pi(\theta)$ and is sampled directly by the MCMC elements.

\begin{table}[t!]
\caption{MCMC parameters and priors for the CP asymmetries, branching ratios and $m_{\rm DM}$.  All priors
are uniform over the indicated range.\label{tab:prior2}}
\begin{center}
\lineup
\begin{tabular}{ll}
\br
 MCMC parameter & Prior \\
\mr
 $\log(m_{\rm DM}/{\rm GeV})$  & $0  \to 5$\\
 $\log(\epsilon_L)$ & $-9 \to 0$ \\ 
 $\log(\epsilon_{\rm DM})$ &  $-9 \to 0$\\
 $\log(B_L)$ &  $-5 \to  0$\\
 $\log(B_{\rm DM})$ & $-5 \to 0$ \\
\br
\end{tabular}
\end{center}
\end{table}

The likelihood function $\mathcal{L}(X|\theta)$ denotes the probability of the data $X$ given some theoretical prediction $\theta$ and 
plays a central role in Bayesian inference. Both the abundances of dark matter $\Omega_{\rm DM}$ and baryonic matter $\Omega_{b}$ are variables normally distributed around their mean values given by WMAP measurements~\cite{Komatsu:2010fb}: $\Omega_{\rm DM} \equiv \bar{\Omega}_{\rm DM}\pm\sigma_{\rm DM} = 0.227 \pm 0.014$ and $\Omega_b \equiv \bar{\Omega}_b \pm \sigma_b = 0.0456 \pm 0.0016$. Since we are interested in the ratio of the dark to baryonic matter, Eq.~\ref{eq:IMP}, we define the likelihood as the probability distribution of the model parameter to satisfy that ratio. The likelihood is therefore well described by the so-called ratio distribution~\cite{HINKLEY01121969}, which is constructed as the distribution of the ratio of variables normally distributed with non zero mean. Calling $r =\Omega_{\rm DM}/\Omega_b \equiv f(m_{\rm DM},\epsilon_i,B_i)$, the likelihood reads:
\begin{equation}\label{eq:loglkBE}
\fl {\cal L}_{\rm ratio} = \frac{1}{\sqrt{2 \pi}\sigma_b \sigma_{\rm DM}}  \frac{b(r) c(r)}{a(r)^3} \Phi\Big(\frac{b(r)}{a(r)}\Big) + \frac{\exp\Big(-1/2 (\bar{\Omega}_{\rm DM}^2/\sigma_{\rm DM}^2 +\bar{\Omega}_b^2/\sigma_b^2)\Big)}{a(r)^2 \pi \sigma_b \sigma_{\rm DM}}\,,
\end{equation}
with:
\begin{eqnarray}
a(r) & = & \sqrt{\frac{z^2}{\sigma_{\rm DM}^2}+\frac{1}{\sigma_b^2}}\,,\nonumber\\
b(r) & = & \frac{\bar{\Omega}_{\rm DM}}{\sigma_{\rm DM}^2} r +\frac{\bar{\Omega}_{b}}{\sigma_b^2}\,,\nonumber\\
c(r) & = & \exp\Big(\frac{1}{2}\frac{b(r)^2}{a(r)^2} - \frac{1}{2}(\frac{\bar{\Omega}_{\rm DM}^2}{\sigma_{\rm DM}^2}+\frac{\bar{\Omega}_b^2}{\sigma_b^2})\Big)\,,\nonumber\\
\Phi(u) & = & {\rm Erf}\Big(\frac{u}{\sqrt{2}}\Big)\,.
\end{eqnarray}
with Erf being the error function. The shape of the likelihood function is depicted in figure~\ref{fig:lkhratio}, where it can be seen that the peak is at around $r \sim 5$ as it is expected. In the right plot we show that the ratio distribution is slightly skewed with respect to a gaussian distribution (gray dashed line) with the same variance.

\begin{figure}[t]
\begin{minipage}[t]{0.5\textwidth}
\centering
\includegraphics[width=1.\columnwidth]{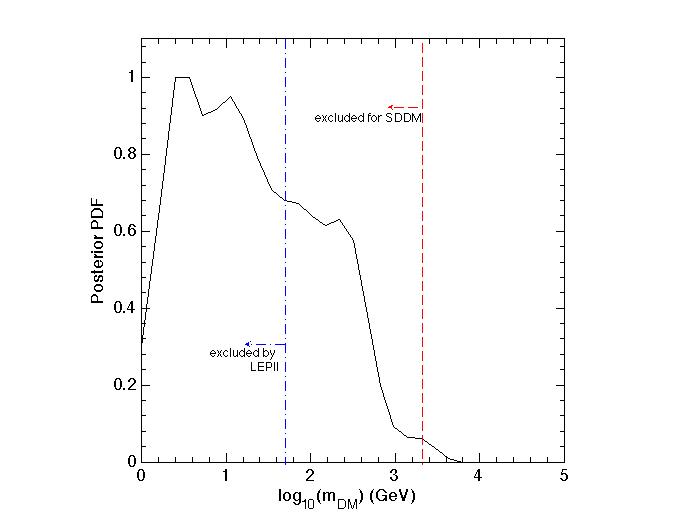}
\end{minipage}
\hspace*{-0.2cm}
\begin{minipage}[t]{0.5\textwidth}
\centering
\includegraphics[width=1.\columnwidth]{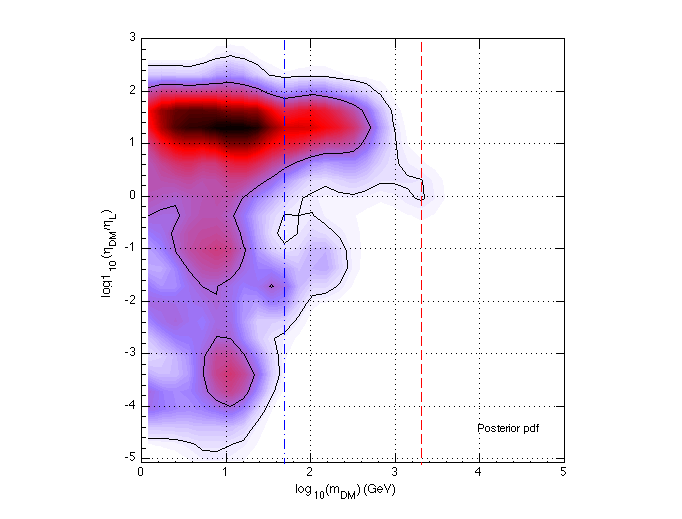}
\end{minipage}
\caption{{\it Left:} 1D posterior pdf for the DM mass $m_{\rm DM}$. {\it Right:} 2D credible regions at 68\% and 95\% C.L. in the \{$m_{\rm DM},\eta_{\rm DM}/\eta_{L}$\}-plane. The vertical dot-dashed blue line denotes the bound from LEPII, while the red dashed vertical line marks the bound from $\chi_0-\bar{\chi}_0$ oscillations for SDDM. All of other parameters in each plane have been marginalized over.}
\label{fig:1DmDM}
\end{figure}

In addition the baryon asymmetry should satisfy the constraints from WMAP: 
\begin{equation}
\label{eq:barasym}
\eta_b = \Big(\frac{n_b}{n_\gamma}\Big)\Big|_0 = \bar{\eta}_b \pm \sigma_{\eta b} = (6.15 \pm 0.25) \times 10^{-10} \,,
\end{equation}
where $\eta_b=7.02 \times \mathcal{S}_{\rm DM} Y_L$, and the density of photon and baryons are computed at present time. The baryon asymmetry is described by a gaussian distribution:
\begin{equation}
\label{eq:lkhb}
\mathcal{L}_L \propto \exp\Big(-\frac{(\eta_b -\bar{\eta}_b)^2}{2 \sigma^2_{\eta b}}\Big)\,.
\end{equation}
Summarizing, the logarithm of the total likelihood is given by the sum of Eqs.~\ref{eq:loglkBE} and~\ref{eq:lkhb}:
\begin{equation}
\ln\mathcal{L}_{\rm asym} = \ln\mathcal{L}_{\rm ratio} + \ln\mathcal{L}_L\,.
\end{equation}

\begin{figure}[t!]
\begin{minipage}[t]{0.5\textwidth}
\centering
\includegraphics[width=1.2\columnwidth]{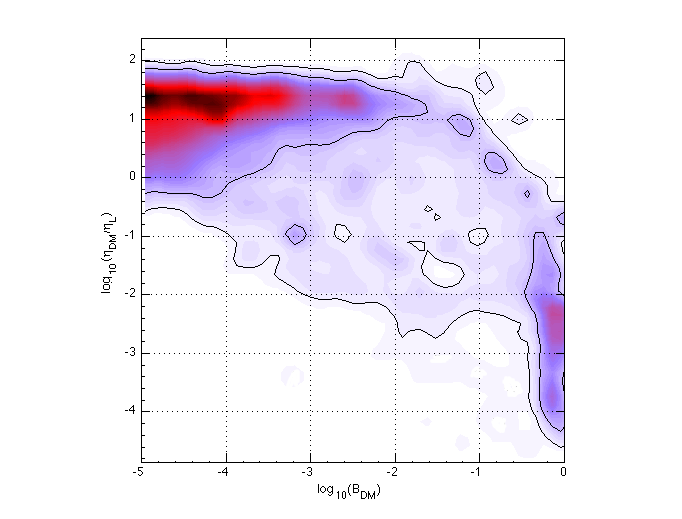}
\end{minipage}
\hspace*{-0.2cm}
\begin{minipage}[t]{0.5\textwidth}
\centering
\includegraphics[width=1.2\columnwidth]{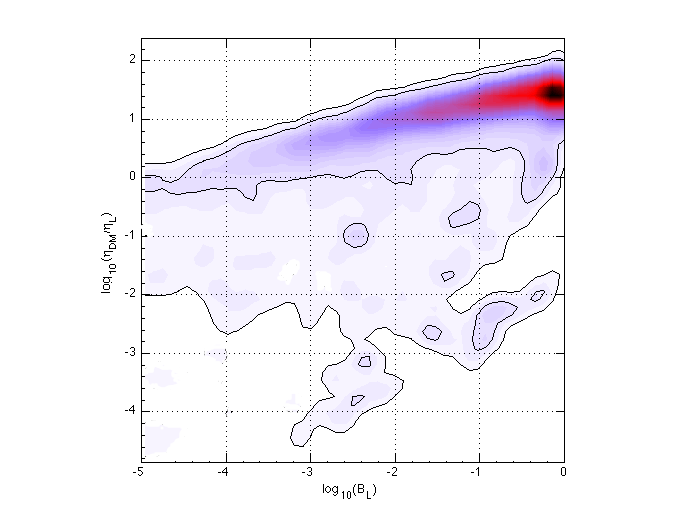}
\end{minipage}
\caption{ {\it Left:} 2D posterior pdf in the \{$B_{\rm DM},\eta_{\rm DM}/\eta_{L}$\}-plane. {\it Right:} Same as left in the \{$B_{\rm L},\eta_{\rm DM}/\eta_L$\}-plane. The credible regions are given at $68\%$ and $95\%$ C.L.. All of other parameters in each plane have been marginalized over.
\label{fig:set2}}
\end{figure}

MCMC techniques require a prior assumption on the probability distribution of the parameters, namely on $\pi(\theta)$. In the absence of theoretical constraints on the model parameter there are a priori no constraints on these quantities. We therefore focus on the regions singled out by successful triplet leptogenesis in~\cite{Hambye:2005tk,Chun:2006sp}, that is values of the CP asymmetries ranging from $10^{-9} - 1$ and branching ratios from 1 to $10^{-5}$. In order not to support a particular scale we choose flat prior on the log distribution for each parameters, as described in table~\ref{tab:prior2}. The dark matter mass is let free to vary between 1 GeV up to 10 TeV (even though masses below 50 GeV are excluded by LEPII). As it is known, if the data are not informative enough a dependence on the prior choice is left in the posterior pdf.

We present in the following the results of the bayesian inference for the case of the asymmetric inert scalar doublet dark matter (SDDM).

In the left panel of figure~\ref{fig:1DmDM} we show the 1D posterior pdf for $m_{\rm DM}$, while all other parameters are marginalized over. We see that all the mass range from 1 GeV up to $\sim 4$ TeV can lead to successful leptogenesis, namely $Y_L \sim 10^{-10}$ and an asymmetric dark matter satisfying the ratio Eq.~\ref{eq:IMP}. The vertical blue line denotes the bound from the $Z$ decay width and masses below 50 GeV are excluded, while the region on the left of the red vertical line is excluded by $\chi_0-\bar{\chi}_0$ oscillations. As one can see from the posterior pdf the most favoured region is at $m_{\rm DM} \sim 10$ GeV, while there are candidates at 100 GeV with smaller statistical significance. With even less probability but still viable are candidates at TeV, which is the range of interest for SDDM. On the right panel of figure~\ref{fig:1DmDM} the 68\% and 95\% credible region are shown in the \{$m_{\rm DM},\eta_{\rm DM}/\eta_{L}$\}-plane. From there we see that 
for DM mass up to around 500 GeV, the preferred values of the ratio $\eta_{\rm DM}/\eta_L$ remains constant to be around 10-50 as these are easily compensated by the small CP asymmetry ratio $\epsilon_{\rm DM}/\epsilon_L$. However, for DM masses above ${\cal O}$(TeV),  $\epsilon_{\rm DM}/\epsilon_L$ is not sufficiently small to compensate with large $\eta_{\rm DM}/\eta_L$ and therefore, the preferred 
values of $\eta_{\rm DM}/\eta_L$ remains to be around unity. Alternatively for a given DM mass, 
smaller values of $\eta_{\rm DM}/\eta_L$ are allowed at 95\% C.L. for $\epsilon_{\rm DM}/\epsilon_L > 1$. 

\begin{figure}[t!]
\begin{minipage}[t]{0.33\textwidth}
\centering
\includegraphics[width=1.2\columnwidth]{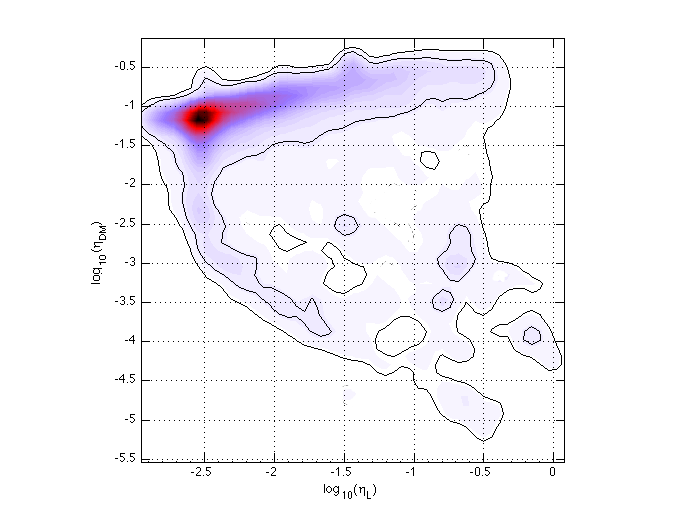}
\end{minipage}
\begin{minipage}[t]{0.33\textwidth}
\centering
\includegraphics[width=1.2\columnwidth]{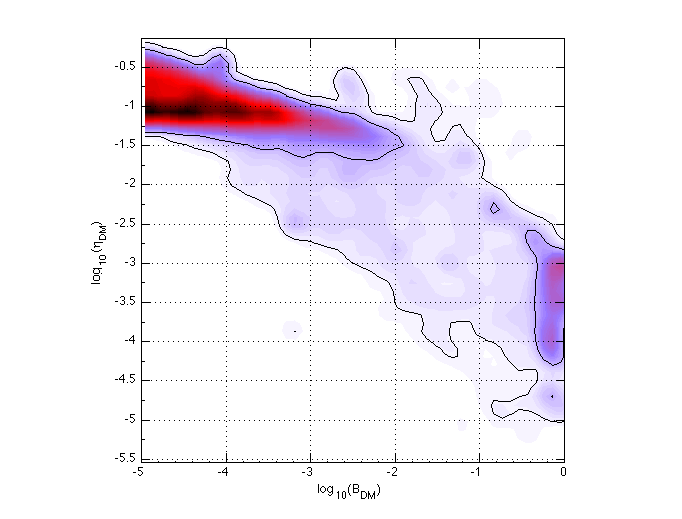}
\end{minipage}
\hspace*{-0.2cm}
\begin{minipage}[t]{0.33\textwidth}
\centering
\includegraphics[width=1.2\columnwidth]{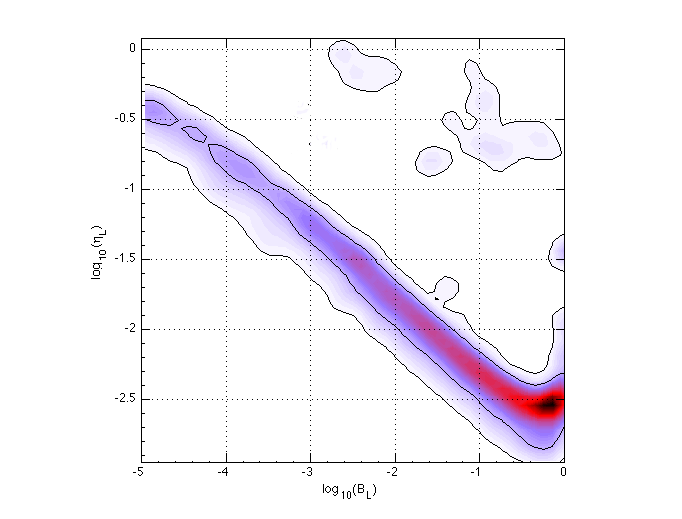}
\end{minipage}
\caption{{\it Left:} 2D posterior pdf in the \{$\eta_{\rm DM},\eta_{L}$\}-plane. {\it Central:}  2D posterior pdf in the \{$B_{\rm DM},\eta_{\rm DM}$\}-plane. {\it Right:} same as left in the \{$B_{L},\eta_{L}$\}-plane. The credible regions are given at $68\%$ and $90\%$ C.L.. All of other parameters in each plane have been marginalized over.
\label{fig:set3}}
\end{figure}

The above preferred values of the parameter space can be understood from the Boltzmann equation 
\ref{boltzman-3} in which the input parameters are the CP asymmetries and the branching ratios. 
Having chosen a mass of the scalar triplet of $10^{10}$ GeV the dominant processes that regulate the boltzmann equations are the decay and inverse decay, and fundamental quantities are the branching ratios. In figure~\ref{fig:set2} we show the correlation of $\eta_{\rm DM}/\eta_L$ versus $B_{\rm DM}$ and $B_L$ respectively in the left and right panels, within the 68\% and 95 \% credible regions. We see that large efficiency ratio $\eta_{\rm DM}/\eta_L$ 
is preferred when $B_L \to 1$ and small $B_{\rm DM} \to 10^{-5}$. This is because larger the value of $B_L$ 
(which implies smaller is the $B_{\rm DM}$ as $\sum_i B_i=1$  with i=$L,H$,DM) the larger is the washout 
due to inverse decay and hence leads to small $\eta_L$. On the contrary smaller is the $B_{\rm DM}$ the washout effect is small due to inverse decay and hence large $\eta_{\rm DM}$. Note that in either case the production of asymmetry is proportional to $\Gamma_1 \propto 1/\sqrt{B_L B_H}$. Therefore when 
$B_L$ approaches towards $10^{-5}$ the asymmetry ($Y_L$) as well as the efficiency ($\eta_L$) get 
increased. On the other hand when $B_{\rm DM}$ approaches towards $1$, which implies small $B_L$, the asymmetry $Y_{\rm DM}$ gets increased but efficiency gets decreased. These behaviours 
of $\eta_{\rm DM}$ and $\eta_L$ can be confirmed from figure~\ref{fig:set3} where we have shown the 2D credible regions at 68\% and 95\% C.L.. The extreme left one, which constitutes the summary of middle and 
right ones, reveals that a successful asymmetric dark matter and lepton asymmetry can be generated 
with small $\eta_L$ and large $\eta_{\rm DM}$. In other words, large $B_L$ and small $B_{\rm DM}$ are 
required in favour of the observed BAU and asymmetric dark matter.  

For sake of reference we report the preferred values of the input CP asymmetries. The preferred values range between $10^{-9}-10^{-2}$ for $\epsilon_{\rm DM}$, respectively to $B_{\rm DM}=10^{-5}-0.5$. A more tighter range is selected in the case of the lepton CP asymmetry: $10^{-8}-10^{-5}$  again for $B_L$ ranging from its extremal values. We remark in addition that for large $B_{\rm DM}$ and small $B_L$ and masses around 50 GeV, the CP asymmetry in the DM sector can be be larger by an order of magnitude with respect to $\epsilon_L$ to compensate the small value of $\eta_{\rm DM}/\eta_L$.

Regarding the inert fermionic doublet DM (FDDM) candidate, the discussion is very similar. We have verified that there are no substantial differences in the selected 1D and 2D credible regions as the Boltzmann 
equations are same in both cases. A small difference comes from the internal degrees of freedom which makes the equilibrium value of a FDDM different from a SDDM. Therefore, in case of FDDM, the allowed mass range goes up to a few TeV starting from 50 GeV as shown in figure~\ref{fig:1DmDM}.

\begin{figure}[t]
\begin{minipage}[t]{0.5\textwidth}
\centering
\includegraphics[width=1.\columnwidth]{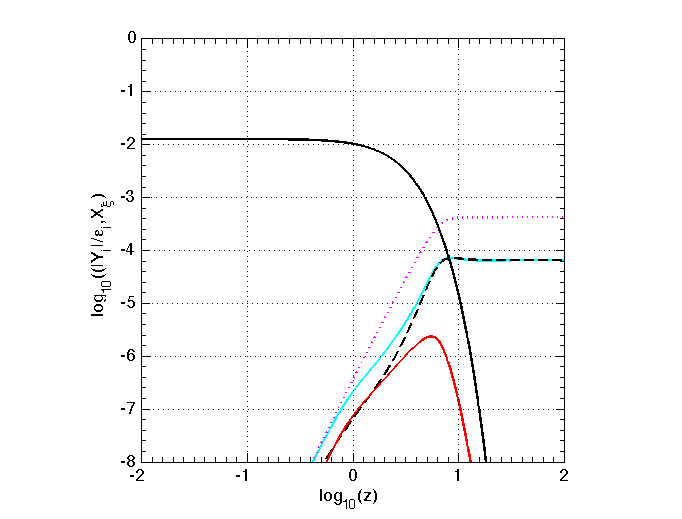}
\end{minipage}
\hspace*{-0.2cm}
\begin{minipage}[t]{0.5\textwidth}
\centering
\includegraphics[width=1.\columnwidth]{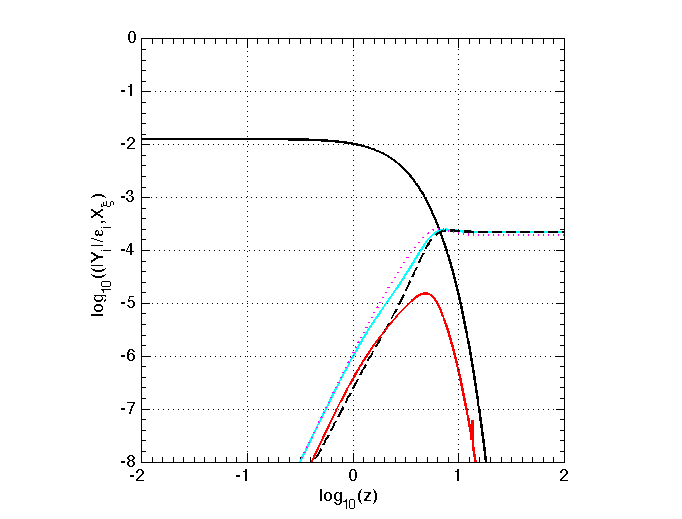}
\end{minipage}
\caption{{\it Left:} Absolute value for the Yield of leptons (cyan solid), DM (dotted magenta), Higgs (dashed black), $\xi$ asymmetry (solid red) plus scalar triplet abundancy (black solid), for a successful point with $m_{\rm  DM}=86$ GeV, $B_L=0.09$, $B_{\rm DM}=4.1 \times10^{-4}$, $\epsilon_{L}=2.6 \times 10^{-6}$, $\epsilon_{\rm DM}=1.1\times  10^{-8}$ which leads to $r\equiv \Omega_{\rm DM}/\Omega_b=4.75$, $Y_L=1.7 \times 10^{-10}$ and $\eta_{\rm DM}/\eta_L=6.53$. {\it Right:} Same as left for $m_{\rm DM}=2$ TeV, $B_L=9.5 \times 10^{-3}$, $B_{\rm DM}=2.6 \times 10^{-5}$, $\epsilon_L=7 \times 10^{-7}$, $\epsilon_{\rm DM}=1.2 \times 10^{-9}$, $\Omega_{\rm DM}/\Omega_b=5.4$ and $Y_L=1.6\times 10^{-10}$ and $\eta_{\rm DM}/\eta_L=0.86$. The $|Y_i|$ are rescaled in terms of CP asymmetries.}
\label{fig:points}
\end{figure}

In figure~\ref{fig:points} we show the behavior of the Yields for leptons, Higgs, DM, scalar triplet and $X_\xi$ for two particular points. The first point in the parameter space is shown in the left panel, which leads to a successful model for FDDM with a mass of $\sim 86$ GeV, $r \sim 4.8$ and $Y_L=1.7 \times10^{-10}$. The second point in the parameter space is depicted in the right panel and accounts for a SDDM with $m_{\rm DM}\sim 2$ TeV, $r \sim 5.3$ and successful baryon asymmetry, $Y_L= 1.6\times 10^{-10}$. The details about the parameters are given in the caption. These two points are representative of the behavior discussed above. 
In particular, for the left panel the branching ratios are $B_L=0.09$ and $B_{\rm DM}=4.1\times 10^{-4}$, 
which implies small $\eta_L$ and large $\eta_{\rm DM}$. Therefore, the ratio of $\eta_{\rm DM}/\eta_L$ is 
maximum and can be confirmed from figure~\ref{fig:set2}. For the figure in right panel the branching 
ratios are both small $B_L=9 \times 10^{-3}$ and $B_{\rm DM}=3 \times 10^{-5}$, which implies $\eta_{\rm DM}$ and $\eta_L$ are comparable. As a result the ratio $\eta_{\rm DM}/\eta_L\sim 0.9$ and the large DM mass is compensated by the very small CP asymmetry ratio. This behavior can be confirmed from 
figure~\ref{fig:set2}.

\begin{figure}[t]
\begin{minipage}[t]{0.33\textwidth}
\centering
\includegraphics[width=1.2\columnwidth]{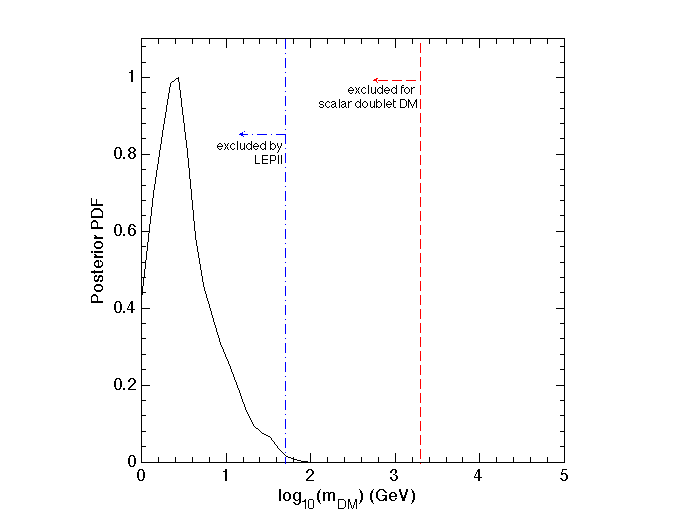}
\end{minipage}
\hspace*{-0.2cm}
\begin{minipage}[t]{0.33\textwidth}
\centering
\includegraphics[width=1.2\columnwidth]{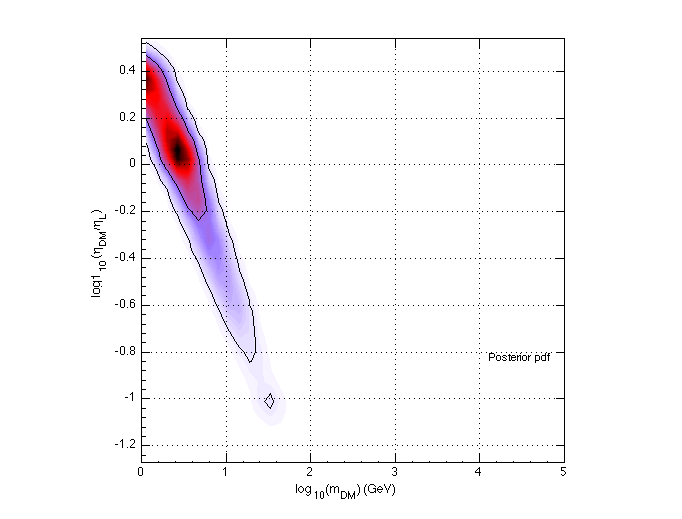}
\end{minipage}
\hspace*{-0.2cm}
\begin{minipage}[t]{0.33\textwidth}
\centering
\includegraphics[width=1.2\columnwidth]{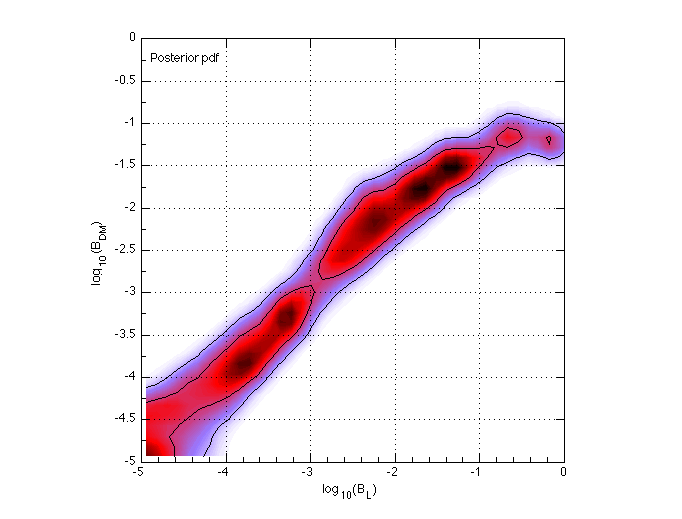}
\end{minipage}
\caption{{\it Left:} 1D posterior pdf (black solid line) for the DM mass $m_{\rm DM}$. The vertical dot-dashed blue line denotes the bound from LEPII. {\it Middle:} 2D credible regions at 68\% and 95\% C.L. in the \{$m_{\rm DM},\eta_{\rm DM}/\eta_L$\}-plane. {\it Right:} Same as middle in the \{$B_{\rm DM},B_{L}$\}-plane. All of other parameters in each plane have been marginalized over.}
\label{fig:eqcp}
\end{figure}

For checking the consistencies, we have investigated the behavior of efficiency factors in case of equal CP asymmetries in DM and lepton channels. There are two interesting results that come out from the MCMC run, shown in figure~\ref{fig:eqcp}. In the left panel the 1D posterior pdf for the DM mass is depicted. We note that equal asymmetries lead to a upper bound to the asymmetric dark matter mass of $\mathcal{O}(50)$ GeV 
and can be applied to the case of FDDM. This is some how expected because the ratio of CP asymmetries can not compensate the increasing DM mass and therefore around 100 GeV there are no more candidate which can fulfill the DM to baryon requirement. This can be seen from the central panel where the 2D credible regions in the \{$m_{\rm DM},\eta_{\rm DM}/\eta_L$\}-plane are shown: the ratio of efficiency drops very rapidly as soon the DM mass increases. As a consequence we found the known results of a light DM mass, around 10 GeV, being favoured. In the right panel the 2D credible regions are shown in the $B_L-B_{\rm DM}$-plane. We see that the preferred branching ratios are similar in magnitude as expected and therefore the efficiencies are comparable. In other words the preferred ratio $\eta_{\rm DM}/\eta_L \gwig 0.1$ for $m_{\rm DM} \lwig 50$ GeV.

\section{Inelastic scattering and Asymmetric DM}\label{sec:inel}

In this section we briefly recall the definition of event rate in a detector and discuss the features of inelastic scattering. For each experiments we describe the 
likelihood functions used in the data analysis and the choice of priors. In the case of direct detection signals, the only free parameters of the model are $m_{\rm DM}$ and the mass splitting $\delta$. For details on the experimental set up and bayesian inference we refers to~\ref{app1} and~\cite{Arina:2011si}.

\subsection{Experiment description, likelihoods and priors}\label{sec:intheo}
The direct detection experiments aim to detect or set limits on nuclear recoils arising from the scattering 
of DM particles off target nuclei. The energy $E_R$ transferred during the collision between the incident particle 
with mass $m_{\rm DM}$ and the nucleus with mass $M_{\cal N}$  is of the order of the keV for a mean DM velocity 
of $v/c\sim 10^{-3}$ in the Galactic halo. The differential spectrum for such recoils, measured in events 
per day/kg/keV, is given by
\begin{equation}
\label{eq:diffrate}
\frac{\rmd R}{\rmd E_{R}} = \frac{\rho_{\odot}}{m_{\rm DM}} \frac{\rmd\sigma}{\rmd E_{R}} \; \eta (E_R,t)\,,
\end{equation}
where $\rmd \sigma/\rmd E_{R}$ encodes all the particle and nuclear physics factors, $\rho_{\odot} \equiv 
\rho_{\rm DM}(R_{\odot})$ is the DM density at the Sun position and $\eta (E_R,t)$ is the mean inverse velocity 
of the incoming particles that can deposit a given recoil energy $E_R$:
\begin{equation}
\label{eq:eta}
\eta (E_R,t) =  \int_{v_{\rm min}} {\rm d}^3 \vec{v}\  \frac{ {\rm f}(\vec{{v}}(t))}{{v}}  \,,
\end{equation}
where the velocity $\vec{v}$ is taken with respect to the Earth frame. The quantity $v_{\rm min}$ is the minimum 
velocity needed to lead to a recoil inside the detector:
\begin{equation}
v_{min} = c \sqrt{\frac{1}{2 M_{\mathcal N} E_R}} \Big(\frac{M_{\mathcal N} E_R}{\mu_n}+\delta\Big)  \,,
\label{eq:vmin}
\end{equation}
where $\mu_n$ the WIMP-nucleus reduced mass and $M_{\mathcal N}$ is the nucleon mass. $\delta$ denotes the mass 
splitting between the DM particle and the 
excited state, therefore proportional to $\lambda_5$ in case of SDDM and Majorana mass in case of FDDM. The value 
required by preserving the asymmetry, $\lambda_5 \sim 10^{-7}$ leads precisely to $\delta \sim 100$ keV, the right 
order of magnitude for inelastic scattering in case of SDDM. On the other hand, in case of FDDM, $\delta \sim m$, 
the Majorana mass of FDDM.

The total event rate per unit detector mass is obtained integrating in a given energy bin $[E_1,E_2]$ Eq.~(\ref{eq:diffrate}),
\begin{equation}
\label{eq:totrate}
R(t) = \int_{E_1}^{E_2} {d}E_R\  \epsilon(E_R)\   \frac{\rmd R}{\rmd E_R}\, ,
\end{equation}
where $\epsilon$ is the energy dependent efficiency of the detector. The expected number of event observed in a detector is given by:
\begin{equation}
S = M_{\rm det} T R(t)\,,
\label{eq:Ntot}
\end{equation}
where $M_{\det}$ is the detector mass and $T$ is the exposure time. 

For some detectors, like scintillators, the recoiling nucleus may loose energy by collisions with other nuclei, hence in form of heat, or through collisions with electrons, which create scintillation light. The observed energy released in scintillation light (typically expressed in keVee) is related to the nuclei recoil energy through the quenching factor $q$, $E_{\rm scint} = q E_R$.

The particle physics cross-section for coherent inelastic scattering $S \ \mathcal{N} \to A\  \mathcal{N}$, mediated by the Z exchange on $t$-channel, is 
parameterised as:
\begin{eqnarray}
\label{eq:pppart}
\frac{\rmd \sigma}{\rmd E_{R}} & = & \frac{M_{\cal N}}{2 \mu^2_n}\ \frac{G_F^2}{2 \pi  f_n^2} \ \Big( (A-Z)f_n +  Z f_p\Big)^2 F^2(E_R)\nonumber\\
& = &  \frac{M_{\cal N}}{2 \mu^2_n}\ \frac{G_F^2}{2 \pi} \ \Big( A +  Z\  (4 \sin^2\theta_{\rm W}-1)\Big)^2 F^2(E_R) \, ,
\end{eqnarray}
where $Z$ and $A$ are respectively the number of protons and the atomic number of the element, $G_F$ is the Fermi constant and $\sin^2\theta_{\rm W}$ is the Weinberg angle. For $SU(2)_L$ doublets with hypercharge 1/2 the couplings to proton and neutron are different, i.e. $f_n \neq f_p$ in contrast to the 
standard elastic scattering. We note that the cross-section is no longer a free parameter. The nuclei form factor $F^2(E_R)$ characterizes the loss of coherence for non zero momentum transfer and is described as the Helm factor~\cite{Helm:1956zz,Lewin:1995rx}.

Regarding $\eta(E_R,t)$ in equation~\ref{eq:eta}, we consider the velocity distributions generated by a cored isothermal and the NFW~\cite{Navarro:1996gj} density profile marginalized over the astrophysical variables ($v_0,v_{\rm esc}$ and $\rho_\odot$), while the standard maxwellian halo is presented in~\ref{app2} for sake of reference. Given a DM density profile and assuming equilibrium between gravitational attractive force and pressure, a corresponding velocity distribution arises from the Eddigton formula~\cite{Binneybook}. We consider these two DM density profiles because they lead to velocity distributions that differ mainly in the selected mean and escape velocity values~\cite{Arina:2011si}, which are fundamental quantities in inelastic scattering. Indeed the splitting factor $\delta$ in Eq.~\ref{eq:vmin} means that only the very high velocity particles will have enough energy to produce a recoil in the detector. This can be seen re-expressing it in terms of target and DM masses
\begin{equation}
\Big(\frac{v}{c}\Big)^2 > \frac{2 \delta (M_\mathcal{N} + m_{\rm DM})}{m_{\rm DM} M_{\mathcal{N}}}\,.
\end{equation}
In addition experiments with heavy nuclei will have a large sensitivity to the high tail of the velocity distribution. We therefore discuss the DAMA experiment because of the Iodine, CRESST-II on W and Xenon100. The Germanium is more sensitive to particle of mass of the order of 50-70 GeV but we consider the dedicated analysis for inelastic scattering~\cite{Ahmed:2010hw}. We do not consider the CoGeNT~\cite{Aalseth:2011wp} experiment since only very light DM can account for its excess. We do not consider Zeplin-III~\cite{Lebedenko:2008gb} since it has analogous sensitivity than CRESST-II.

\paragraph{CRESST-II}
CRESST is a cryogenic experiment running at the Laboratori Nazionali del Gran Sasso. The 33 detector modules are made by CaWO$_4$ crystal, each of a mass of 333 g. In this analysis we consider the second run of CRESST, carried out in 2007 and in particular the data on Tungsten~\cite{Angloher:2008jj}. These data are obtained with two detector modules, leading to a total exposure of 30.6 kg days on W after cuts, in the energy range of 10-40 keV.  Three events have been seen, which are compatible with the expected background, mainly from neutrons, of $\sim 0.063$ kg days. We therefore use a background value $B=3$ with $\sigma_B = 10\%$ of $B$. The likelihood is described by the poisson probability of seeing three events for a given theoretical prediction $S$ and a given background $B$:
\begin{equation}
\label{eq:cresst}
\ln{\cal L}_{\rm Cresst} (3|S,B)= - (S+B) + 3 \ln \left(S+B \right)\,.
\end{equation} 
The effective likelihood we used in the analysis is marginalized numerically over the background:
\begin{equation}
\label{eq:effcresst}
\ln {\mathcal L}^{\rm eff}_{\rm Cresst} = \int_0^{\infty} \rmd B \ \ln{\mathcal L}_{\rm Cresst}   (3 | S,B)\  p(B)\,, 
\end{equation}
where $p(B)$ is the probability function of the background, modeled as a gaussian distribution. The invariant $90_S\%$ confidence level, based on the $S$-signal, corresponds to $\Delta\chi^2 \sim 3.34$.

\paragraph{CDMS on Germanium}
The CDMS collaboration has published a dedicated analysis for inelastic DM~\cite{Ahmed:2010hw}, which we use for constructing the likelihood of the experiment. The total exposure is 969 kg days and the energy range is from 10 keV up to 150 keV. From $\Delta E_1=10-25$ keV 8 events has been found, with an expected background of $5.88_{-1.75}^{+2.33}$, while in the remaining energy range, $\Delta E_1=25-150$, 3 events survive all the cuts and have an expected background of $0.93_{-0.36}^{+0.58}$. The background accounts for all surface events and cosmogenic particles. For the detector efficiency we used the red dotted curve presented in figure 5 of~\cite{Ahmed:2010hw}. 

The total likelihood is the sum of the contribution from the two energy range $\Delta E_1$ and $\Delta E_2$. Each partial likelihood follows the poisson distribution:
\begin{equation}
\label{eq:cdmsge}
\ln{\cal L}_{\rm CDMS} (11|S,B)= \ln{\cal L}_{\Delta E_1} (8|S,B)+ \ln{\cal L}_{\Delta E_2} (3|S,B) \,,
\end{equation} 
with
\begin{eqnarray}
\ln{\cal L}_{\Delta E_1} (8|S,B) & = & - (S+B)+ 8 \ln \left(S+B \right)\,,\nonumber\\
\ln{\cal L}_{\Delta E_2} (3|S,B) & = & - (S+B) + 3 \ln \left(S+B \right)\,.\
\end{eqnarray}
We then marginalise numerically over the background: 
\begin{equation}
\label{eq:bckm}
{\cal L}^{\rm eff}_{\rm CDMS} = \int_0^{\infty} \rmd B \ {\cal L}_{\rm CDMS}   (11 | S,B)\  p(B), 
\end{equation}
to get the effective likelihood we use in computing the exclusion bound.

The $90_S$\% confidence interval corresponds to $\Delta\chi^2 = 2.5$, obtained considering that in the whole energy range there are 11 events with an expected background of 6.

\paragraph{DAMA}
The DAMA likelihood for the modulated rate is described in~\cite{Arina:2011si} and follows a Gaussian distribution:
\begin{equation}
 \ln\mathcal{L}_{\rm DAMA} = - \sum_{i=1}^{N_{\rm bin}} \frac{(s_i-\bar{s}^{\rm  obs}_i)^2}{2 \sigma_i^2} \,, 
\end{equation}
where $s_i$ and $\bar{s}^{\rm obs }_i$ are the theoretical and the mean observed modulation respectively in the $i$th energy bin, $\sigma_i$ is the associated 
uncertainty in the observed signal. We use in this analysis  the 12-bin data from figure~9 of~\cite{Bernabei:2008yi}. The quenching factors $q_{\rm Na}$ and $q_{\rm I}$ are taken to be free parameters in our analysis, which we vary over their respective allowed range~\cite{Bernabei:1996vj,Chagani:2008in}, as reported in table~\ref{tab:prior1}. In addition we require that the unmodulated predicted signal does not overcome the total unmodulated rate in figure~1 of~\cite{Bernabei:2008yi}, namely in each energy bin the predicted total rate should be at most equal to the measured rate.

\paragraph{Xenon100}
As far as it concerns Xenon100 (Xe100 hereafter) experiment, we use the last data release for inelastic scattering~\cite{Aprile:2011ts}. The likelihood is given by a Poisson distribution for three seen events times a gaussian which takes into account the uncertainties on the scintillation efficiency $L_{\rm eff}$. These latter however affect only the low mass DM region. In addition the uncertainties over the background are marginalized over. For details about this experiment we refer to~\cite{Arina:2011si}.

\begin{table}[t!]
\caption{MCMC parameters and priors for the model parameter space and experimental systematics (nuisance parameters).  All priors
are uniform over the indicated range.\label{tab:prior1}}
\begin{center}
\lineup
\begin{tabular}{lll}
\br
Experiment& MCMC parameter & Prior \\
\mr
All & $\log(m_{\rm DM}/{\rm GeV})$  & $0  \to 5$\\
All & $\delta/{\rm keV}$ & $0 \to 200$ \\ 
DAMA & $q_{\rm Na}$ &  $0.2 \to 0.4$\\
DAMA & $q_{\rm I}$ &  $0.06 \to  0.1$\\
Xenon100 & $L_{\rm eff}$& $-0.01 \to 0.18$ \\
\br
\end{tabular}
\end{center}
\end{table}

\paragraph{Choice of priors and pills of bayesian inference} As for the bayesian inference we follow closely the approach of~\cite{Arina:2011si}. We consider a full bayesian analysis without discussing profile likelihoods: indeed in case of informative data (as for DAMA experiment) the posterior pdf and the profile likelihood are equivalent, while in the case of exclusion bounds, in order to be insensitive on the choice of priors, we use the invariant bound for the $x_S\%$  credible region, as described in~\ref{app1}, which can have a bayesian interpretation in terms of probability for the $S$ signal.

Having specified the likelihood functions for each experiments, the only missing element for Bayes theorem, Eq.~\ref{eq:bayes}, is the choice of priors. As in the previous section, the prior on the DM mass is chosen flat on a logarithmic scale on the same range (note that it is the only parameter in common with the asymmetry generation MCMC), while for $\delta$ we chose a flat prior, since the scale of this parameter is known by the requirement of inelastic scattering. The range is given in table~\ref{tab:prior1}, together with the priors for the systematic parameters in each experiment. 

\paragraph{Astrophysics}
In addition to the candidate mass $m_{\rm DM}$, the mass splitting $\delta$ between the DM and its excited state and the nuisance parameters in the experimental set-ups, two further free parameters are used to characterise 
the DM velocity distribution: the virial mass of the DM halo, and its concentration. These additional parameters are, however, also constrained
by astrophysical observations on the velocity of the local standard at rest, $v_0$, on the escape velocity for the DM halo, $v_{\rm esc}$ and on the DM density at the sun position $\rho_\odot$, which all enter in equations~\ref{eq:diffrate} and~\ref{eq:eta}. The gaussian priors and the astrophysical likelihood are given in details in~\cite{Arina:2011si}. Only in the case of Maxwellian velocity distribution we do not vary the astrophysical observable in their allowed range but keep them fixed at their mean values, which are $\bar{v}_0 = 230 \ {\rm km s^{-1}}$~\cite{Reid:2009nj,Gillessen:2008qv}, $\bar{v}_{\rm esc} = 544\  {\rm km s^{-1}}$~\cite{Smith:2006ym,Dehnen:1997cq} and $\bar{\rho}_\odot = 0.4\  {\rm GeV cm^{-3}}$~\cite{Weber:2009pt,Salucci:2010qr}.

\subsection{Results for scalar and fermionic candidates}\label{sec:Inres}

In this section we present our inference analysis for the considered experiments. Before coming to the results for the scalar and fermionic candidate we would like to make few general comments about inelastic scattering mediated by the $Z$ boson and the DAMA region.

\begin{figure}[t]
\begin{minipage}[t]{0.5\textwidth}
\centering
\includegraphics[width=1.1\columnwidth]{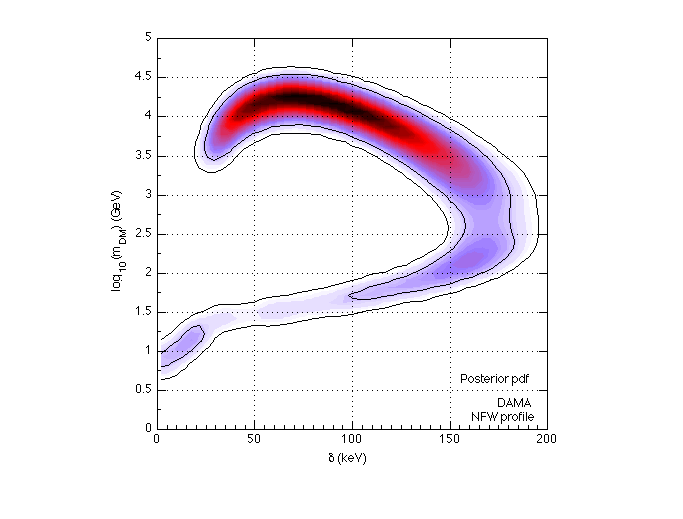}
\end{minipage}
\hspace*{-0.2cm}
\begin{minipage}[t]{0.5\textwidth}
\centering
\includegraphics[width=1.1\columnwidth]{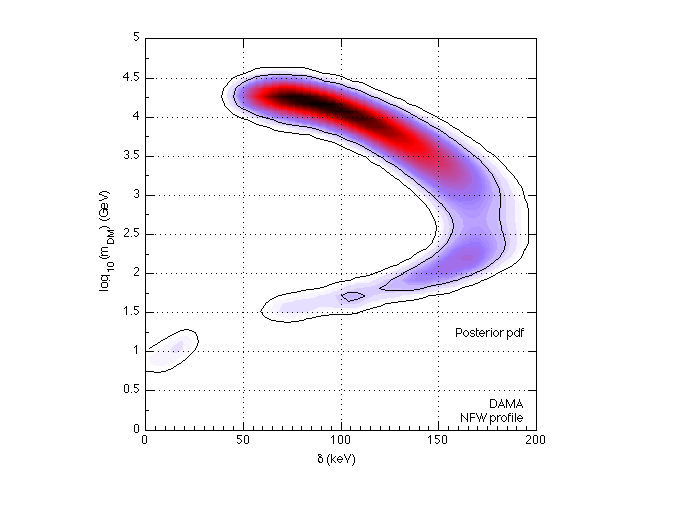}
\end{minipage}
\caption{{\it Left}:  2D marginal posterior for DAMA in the parameters space \{$\delta,m_{\rm DM}$\} and the NFW density profile.
{\it Right}: Same as left with the additional constraint of the total unmodulated rate.
\label{fig:DamaTrNFW}}
\end{figure}

\begin{figure}[t]
\begin{minipage}[t]{0.5\textwidth}
\centering
\includegraphics[width=1.1\columnwidth]{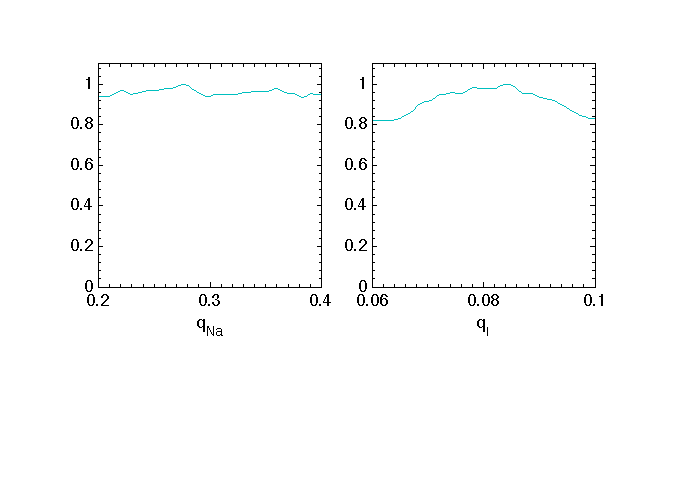}
\end{minipage}
\hspace*{-0.2cm}
\begin{minipage}[t]{0.5\textwidth}
\centering
\includegraphics[width=1.1\columnwidth]{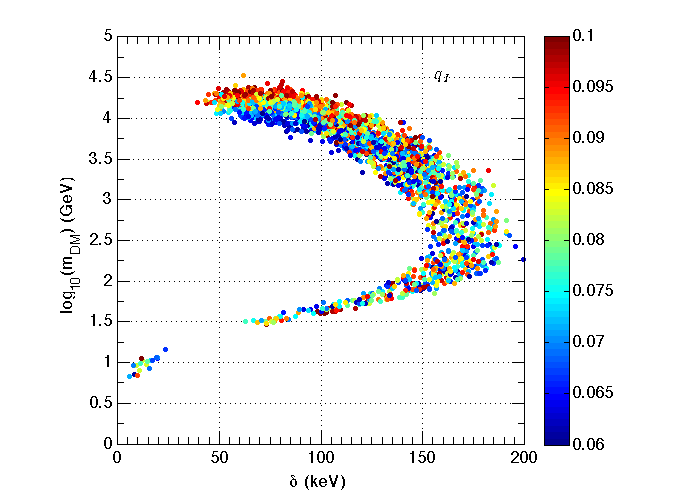}
\end{minipage}
\caption{Inference for DAMA assuming the NFW density profile for the DM halo.
{\it Left}: 1D marginal posterior pdf for the quenching factor of Sodium and Iodine, as labeled. {\it Right}:  3D marginal posterior for \{$\delta,m_{\rm DM},q_{\rm I}$\}, where the $q_{\rm I}$ direction is represented by the colour code.
\label{fig:Damaq}}
\end{figure}

Figure~\ref{fig:DamaTrNFW} shows the preferred DAMA region for inelastic scattering in the plane $\{\delta,m_{\rm DM}\}$ for a NFW density profile. In the left plot only the modulated signal is considered, while on the right-hand plot we add the additional constraint on the total rate: it follows that the region with small $\delta \sim 20$ keV and large masses, $\mathcal{O}(10)$ TeV, is excluded. In addition the mass range around $50$ GeV and small mass splitting is disfavoured, while an island at lower masses and splittings survives. The same behavior is retrieved for the cored isothermal halo. In figure~\ref{fig:Damaq} the dependence on the quenching factors is depicted. As expected $q_{\rm Na}$ is unconstrained, as shown in the left plot by the flat 1D marginal posterior (cyan solid line) while for $q_I$ one might claim for a slight preference for values around 0.08 although it is statistically insignificant. The quenching factors for inelastic, Z mediated scattering result to be less constraint than the elastic spin-independent case~\cite{Arina:2011si}. The right panel illustrates the correlation between  \{$\delta,m_{\rm DM}$\} and the quenching factor on Iodine. There is a clear  dependence on $q_{\rm I}$ for masses between 3 and 30 TeV and splittings in the range 50-100 keV: smaller value of the quenching factor favours smaller splitting and lighter masses. All the remaining region does not show a correlation between the model parameters and $q_{\rm I}$. The small island at masses of few GeV is due to scattering on Sodium and therefore correlated to $q_{\rm Na}$.

\begin{figure}[t]
\begin{minipage}[t]{0.33\textwidth}
\centering
\includegraphics[width=1.1\columnwidth]{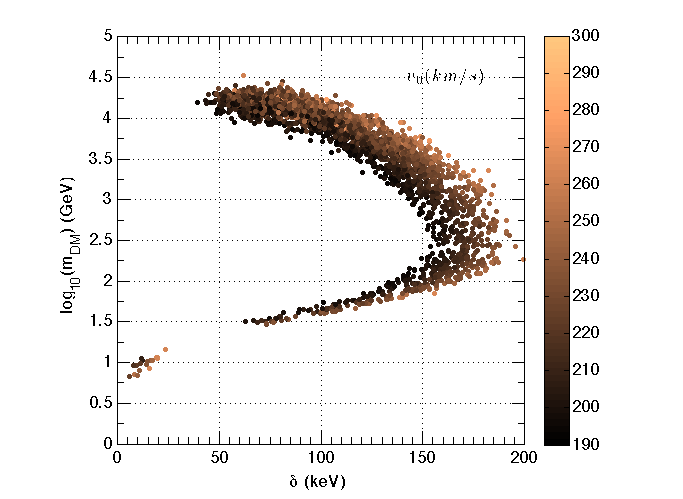}
\end{minipage}
\hspace*{-0.2cm}
\begin{minipage}[t]{0.33\textwidth}
\centering
\includegraphics[width=1.1\columnwidth]{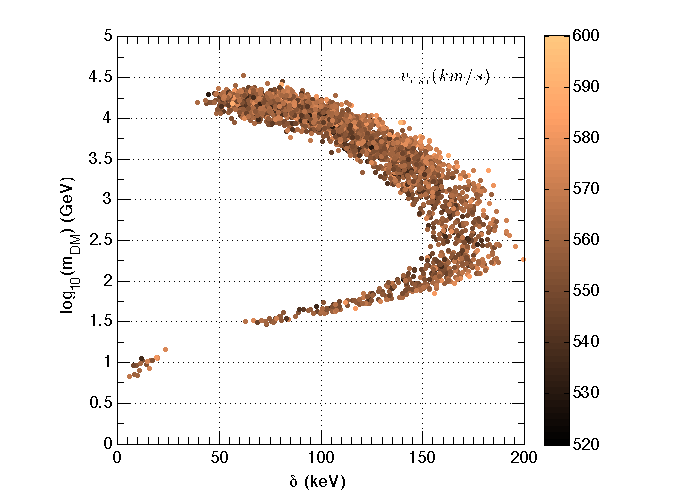}
\end{minipage}
\hspace*{-0.2cm}
\begin{minipage}[t]{0.33\textwidth}
\centering
\includegraphics[width=1.1\columnwidth]{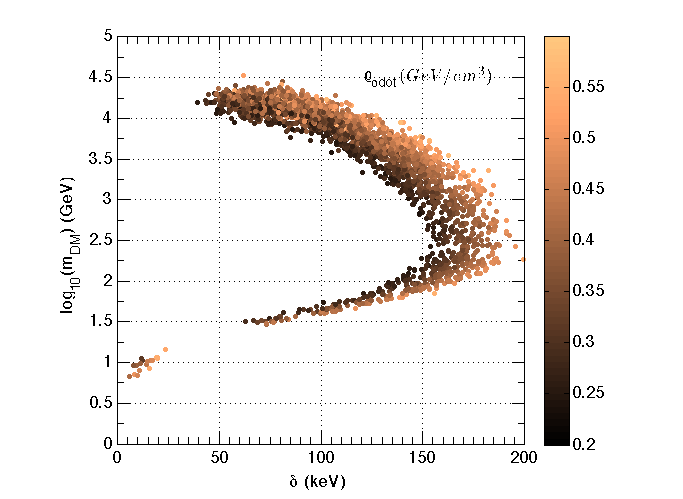}
\end{minipage}
\caption{Inference for DAMA assuming the NFW density profile for the DM halo.
{\it Left}:  3D marginal posterior for \{$\delta,m_{\rm DM},v_0$\}, where the $v_0$ direction is represented by the colour code. {\it Center, right}: Same as left, but for \{$\delta,m_{\rm DM},v_{\rm esc}$\} and \{$\delta,m_{\rm DM},\rho_{\odot}$\} respectively.
\label{fig:DamaNFWq}}
\end{figure}

The dependence on the astrophysical observables and NFW density profile for inelastic scattering is shown in figure~\ref{fig:DamaNFWq} with the 3D marginal posteriors for \{$\delta, m_{\rm DM}$\} and a third parameter direction  $v_0$, $v_{\rm esc}$ and $\rho_\odot$. The DAMA signal favours the high tail of the velocity distribution from the central panel, where the larger values of $v_{\rm esc}$ are preferred. From the left and right panel we see that in the `croissant'-shaped region the internal parts are due to circular velocity below $\bar{v}_0$ and DM density close to $0.2\  {\rm GeV cm^{-3}}$. The increase of $v_0$ and $\rho_\odot$ favours instead the outer parts of the region, in example very large mass splitting $\sim 190$ keV and masses around the TeV scale.

\begin{table}[t!]
\small
\caption{1D posterior modes and $90\%$ credible intervals for the circular velocity $v_0$,
escape velocity $v_{\rm esc}$, and the local DM density $\rho_{\odot}$ for DM density profiles considered in this work. \label{tab:allparams}}
\begin{center}
\lineup
\begin{tabular}{ l | lll }
\br
& $v_0$ (${\rm  km \ s}^{-1}$) & $v_{\rm esc}$ (${\rm km \ s}^{-1}$) & $\rho_{\odot}$  ($ {\rm GeV \ cm}^{-3}$) \\ 
\br
\bfseries{Cored Isothermal} & & &\\
DAMA & $211^{+25}_{-18}$ & $629_{-20}^{+22}$ &$0.31_{-0.03}^{+0.05}$ \\
CDMS & $211_{-19}^{+26}$ & $629_{-21}^{+22}$ & $0.31 \pm 0.04$  \\
Xe100 &$210_{-19}^{+27}$ &$628_{-19}^{+25}$ &$0.31_{-0.03}^{+0.05}$  \\
CRESST &$210_{-18}^{+27}$ &$628_{-20}^{+23}$ & $0.31_{-0.04}^{+0.05}$ \\
\mr
{\bfseries NFW} & & & \\
DAMA & $221^{+40}_{-23}$ & $558_{-18}^{+20}$ & $0.38_{-0.10}^{+0.16}$ \\
CDMS & $220_{-21}^{+39}$ & $558_{-16}^{+19} $ & $0.38_{-0.10}^{+0.14}$ \\
Xe100 &$221_{-22}^{+39}$ &$557_{-21}^{+25}$ & $0.38_{-0.11}^{+0.14}$ \\
CRESST &$220_{-21}^{+42}$ &$558_{-17}^{+21}$ & $0.38_{-0.10}^{+0.16}$ \\
\br
\end{tabular}
\end{center}
\end{table}

\begin{figure}
\begin{minipage}[t]{0.5\textwidth}
\centering
\includegraphics[width=1.1\columnwidth]{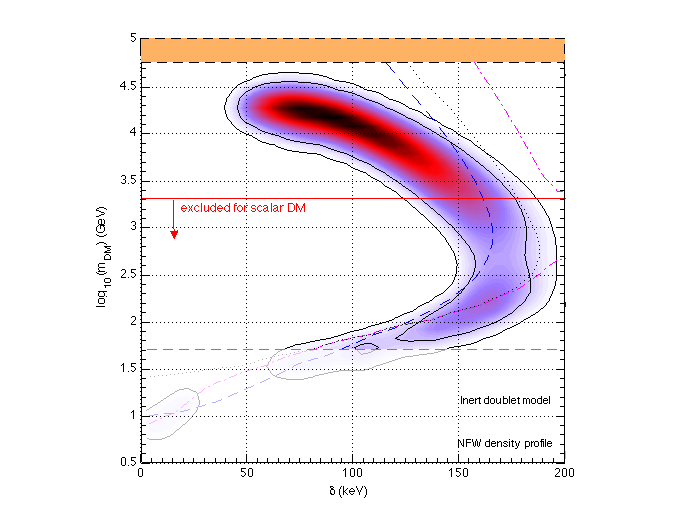}
\end{minipage}
\hspace*{-0.2cm}
\begin{minipage}[t]{0.5\textwidth}
\centering
\includegraphics[width=1.1\columnwidth]{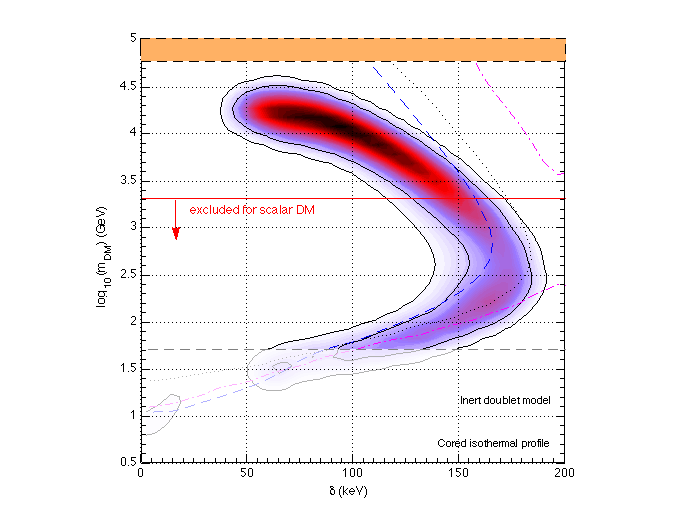}
\end{minipage}
\caption{{\it Left}: 2D credible regions for the individual experimental bounds and regions assuming the NFW DM density profile and marginalizing over the astrophysical uncertainties, combined in a single plot. For DAMA (shaded)  we show the 90\% and 99\% contours.
The $90_S\%$ contours are given by respectively the pink dot-dash curve for Xe100, the dashed blue line for CDMS and the dark green one denotes CRESST. The region below $M_\chi =$ 2 TeV is excluded by $\chi_0-\bar{\chi}_0$ oscillation (horizontal red solid line), while the orange/dark grey region is excluded by unitarity bound and below the dashed gray line by the LEP constraints on the Z decay width. {\it Right}: same as left but for the isothermal cored DM density profile. 
\label{fig:Sall}}
\end{figure}

An analogous behavior holds for velocity distributions arising from the cored isothermal halo. The main difference is that this latter prefers in particular the very high end of the observationally allowed escape velocities. In table~\ref{tab:allparams} the preferred values for the astrophysical observables are indicated for both the NFW dark matter profile and the cored isothermal one. We underline that the difference in the preferred values of $v_{\rm esc}$ will play a role in case of inelastic scattering, even if the statistical significance in the difference of the preferred $v_{\rm esc}$ values is small. Indeed in figure~\ref{fig:Sall} we show all the experimental constraints and the DAMA region in a single plot, on the left for NFW profile and on the right for isothermal cored DM density profile. Firstly we note that the NFW profile favours larger splitting for fitting DAMA with respect to the isothermal profile and secondly the low mass region is larger. The exclusion limits for CDMS (blue dashed), Xe100 (pink dot-dashed) and CRESST (black dotted) are $90_S\%$ confidence intervals and all the region on the left of the curve is excluded. As a general remark the NFW prefers smaller $v_{\rm esc}$, namely the tail of the velocity distribution is constituted by less high speed particles than the isothermal one: there is less room for the detectors to be sensitive to inelastic scattering and therefore the exclusion limits are less constraining. Regarding the NFW profile, up to masses of 80 GeV the most constraining upper bounds is CDMS, then leaving the place to Xe100 that excluded all DM masses above 316 GeV. The trend for the isothermal profile is the same, except that CDMS and Xe100 intercepts at 50 GeV and all masses above 251 are excluded by Xe100. The transparent region below 50 GeV is excluded by LEPII constraint on the $Z$ decay width, while the orange region above 56 TeV is excluded requiring unitarity of the S matrix~\cite{Griest:1989wd}. The CRESST upper bound is comparable to the Xe100 one up to masses of 300 GeV, because even though it has a much smaller total exposure the W is heavier than Xe. For larger masses the effect of the larger total exposure of Xe100 dominates.

In terms of SDDM, an asymmetric candidate is completely excluded: the parameter space that survives the $\chi_0-\bar{\chi}_0$ oscillation bound is severely disfavoured by Xe100, for both DM density profiles. In case of the inert doublet model as standard thermal relic there is still room up from 45 GeV to 300 GeV for a NFW profile and 250 GeV for an isothermal halo. It has been shown that this mass range provides the correct WMAP relic abundance thanks to three body annihilation channels~\cite{Honorez:2010re,LopezHonorez:2010tb} and is in the reach of LHC~\cite{Dolle:2009ft,Miao:2010rg}. On the contrary, the asymmetric fermionic doublet is a good DM candidate up to 200 or 300 GeV depending on the DM density profile, in addition of satisfying the DM to baryon ratio.

\section{Conclusions}\label{sec:concl}

We proposed a simple extension of the SM model to explain the observed ratio $\Omega_{\rm DM}/\Omega_B
\approx 5$ as given by WMAP. We extended the SM by including two heavy triplet scalars whose partial decay to SM leptons and inert (odd under a $Z_2$ symmetry) doublet scalars ($\chi$), or vector like fermions ($\psi$), could explain a common origin of asymmetric dark matter and visible matter via leptogenesis route. Moreover, the induced vev of the triplets also gave rise to neutrino masses, as required by the oscillation experiments, via the type-II seesaw mechanism. Thus a triple unification of asymmetric dark matter, leptogenesis and neutrino masses could be achieved.

We studied the relevant annihilation and scattering processes that arise in the model. The asymmetry in case of inert scalar ($\chi$) doublet dark matter (SDDM)
gets strongly depleted by the contact annihilation process $\chi \chi \to H H$ mediated via $\lambda_5$
coupling. Therefore, the survival of the asymmetry in case of inert SDDM required $\lambda_5 < 10^{-5}$.
Besides that we showed that $\lambda_5 \sim 10^{-7}$ is required for the annual modulation signal at DAMA
while restoring the asymmetry. On the other hand, the inert fermion ($\psi$) doublet dark matter (FDDM) does
not under go any further depletion of asymmetry in comparison to leptons. A strong constraint arose on the mass scale of inert SDDM from the rapid oscillation between $\chi_0$
and its complex conjugate $\overline{\chi_0}$. Below EW phase transition the fast oscillation between
$\chi_0$ and $\overline{\chi_0}$ depletes the asymmetry strongly. Therefore, the survival of the asymmetry
in case of SDDM led to its mass $M_\chi \gwig 2 \TeV$ so that it freezes out before it begins to oscillate.  On the other hand, in case of inert FDDM, the survival of asymmetry does not depend on its mass apart from the LEP constraint that $M_\psi \gwig M_Z/2$. Hence a ${\mathcal O}(100) \GeV$ dark matter is allowed.

We then numerically solved the relevant Boltzmann equations to estimate the efficiency factors of DM and lepton in either scenarios, for a fixed scalar triplet mass of $M_1=10^{10}$ GeV. The model parameter space has been systematically investigated via MCMC techniques. We have singled out the preferred regions in the parameter space that lead to a successful leptogenesis and to an asymmetric DM, namely satisfying $\Omega_{\rm DM}/\Omega_b$ and $n_b/n_\gamma$ ratios. We showed that: 
\begin{enumerate}
 \item 
dark matter, irrespective of SDDM or FDDM, masses up to $\mathcal{O}(\TeV)$ can fulfill the requirement of $\Omega_{\rm DM}/\Omega_b$,
\item 
for observed BAU and asymmetric dark matter large $B_L$ and small $B_{\rm DM}$ are preferred. In particular 
for $B_L\to 1$ and $B_{\rm DM} \to 10^{-5}$ the efficiency ratio $\eta_{\rm DM}/\eta_L$ approaches its maximum value.
\end{enumerate}

The survival of asymmetry in the dark sector leads to inelastic dark matter because the elastic scattering is subdominant in both (SDDM and FDDM) cases. In case of SDDM the small coupling $\lambda_5 \sim 10^{-7} $ gave rise to a mass difference between the excited state and ground state of DM to be ${\mathcal O}(100) \keV$. On the other hand, in case of inert FDDM, the
${\mathcal O}(100) \keV$ mass difference between the ground state and excited state of DM is provided by
its Majorana mass induced by the triplet scalar. By performing a bayesian analysis we found that
an asymmetric SDDM of mass larger than 2 TeV is strongly disfavoured by the Xenon100 data
while an asymmetric FDDM of mass ${\mathcal O}(100) \GeV$ is suitable to explain DAMA annual modulation
signal while passing the latest constraint from Xenon100 experiment.

\ack N.S. would like to thank Jean-Marie Fr\`ere, Thomas Hambye and Michel Tytgat for useful discussions. C.A. acknowledges use of the cosmo computing resources at CP3 
of Louvain University. N.S. is supported by the IISN and the Belgian Science Policy (IAP VI-11).

\appendix

\section{Bayesian inference and MCMC techniques}\label{app1}

In the analysis of the data $X$ the
inference of the posterior probability density as a function of the
parameters, $\mathcal{P} (\theta | X)$, is constructed invoking Bayes' theorem:
\begin{equation}
  \mathcal{P} (\theta | X) {\rm d}\theta \propto\ \mathcal{L}(X |
  \theta) \cdot  \pi(\theta) {\rm d}\theta \,,
\end{equation}
where $\mathcal{L}(X|\theta)$ is the likelihood function and $\pi(\theta)$ denotes the
probability density on the parameter space $\theta$ prior to observing the data $X$.
The posterior pdf represents our state of knowledge about the parameters after taking
into account the information contained in the data, and has an
intuitive and straightforward interpretation in that $\int_V
\mathcal{P} (\theta | X) {\rm d}\theta$ is the probability that the
true value of $\theta$ lies in the volume $V$. 

Since the prior pdf is independent of the data, it needs to be chosen according to one's belief and is thus inherently subjective. 
In the often encountered situation in which no unique theoretically
motivated prior pdf can be derived, one may wish to use one which
does not favour any parameter region in particular.
A common choice is the uniform or top-hat prior
\begin{equation}
\pi_{\rm flat}(\theta)  \propto \left\{ 
    \begin{array}{cl} 
     1, &
     {\rm if}\ \theta_{\rm min} \leq \theta \leq  \theta_{\rm max} ,
     \\ 0, & {\rm otherwise},
    \end{array}
\right.
\end{equation}
if the general order of magnitude of the parameter is known.  Here, the limits $\theta_{\rm min}$ and $\theta_{\rm max}$ should 
be chosen such that they are well beyond the parameter region of interest. If even
the order of magnitude is unknown, one may want to choose a uniform prior in $\log \theta$ space 
instead,
\begin{equation}
\pi_{\rm log}(\log \theta) \ \rmd \log \theta  = \left\{ 
    \begin{array}{cl} 
   \rmd \log \theta, &
      {\rm if} \ \theta_{\rm min} \leq  \theta \leq  \theta_{\rm max},
      \\ 0, & {\rm otherwise},
    \end{array}
\right.
\end{equation}
which 
is equivalent to a $\rmd \theta/\theta$ prior in $\theta$ space. Note that
because the volume element $ \rmd \theta$ is in general not invariant under a parameter
transformation $f: \theta \to \theta'$, a uniform prior pdf on $\theta$ does not yield the same 
probabilities as a uniform prior pdf on $\theta'$ unless the mapping $f$ is linear.  
The same is also true for the posterior probabilities, i.e., $  \mathcal{P} (\theta | X) {\rm d}\theta \neq
  \mathcal{P} (\theta' | X) {\rm d}\theta'$ in general.

While the posterior pdf technically contains all the necessary information for the 
interpretation of  the data, the fact that it is a function in the
$N$-dimensional space of parameters makes it difficult to
visualise if $N>2$.  Being a probability
density, its dimensionality can be easily reduced by
integrating out less interesting ({\it nuisance}) parameter directions
$\psi_i$, yielding an $n$-dimensional marginal posterior pdf,
\begin{equation}
 \label{eq:marg}
 \mathcal{P}_{\rm mar}(\theta_1, ..., \theta_n | X) \propto \int \rmd\psi_1
 ... \rmd\psi_m \ {\cal P}( \theta_1, ..., \theta_n,\psi_1...,
 \psi_m|X) \,, 
\end{equation} 
which is more amenable to visual presentation if $n=1,2$, and can
be used to construct constraints on the remaining parameters.

We employ a modified version of the generic Metropolis--Hastings
sampler~\cite{Metropolis53,Hastings70} included in the public MCMC  code
\texttt{CosmoMC}~\cite{Lewis:2002ah,cosmomc_notes}, to sample the
posterior over the full parameter space. Each point $x_{i+1}$ obtained from a Gaussian random distribution (the so called proposal density) 
around the previous point $x_i$ is accepted to be the next element of the chain with probability:
\begin{equation}
{\rm P}(x_{i+1}) = {\rm min} \Big( 1,\frac{\mathcal{P} (x_{i+1})}{\mathcal{P}(x_i)}\Big)\,.
\end{equation}

The resulting chains are analysed with an adapted version of the accompanying package
  \texttt{GetDist}, supplemented with \texttt{matlab} scripts from the package \texttt{SuperBayeS}~\cite{superbayes,Trotta:2008bp}. 
  One- or two-dimensional marginal posterior pdfs are obtained
from the chains by dividing the relevant parameter subspace into bins and
counting the number of samples per bin.  An $x\%$ credible interval or region containing $x\%$ of the total volume of
$\mathcal{P}_{\rm mar}$ is then constructed by demanding that  $\mathcal{P}_{\rm mar}$  at any point inside the region 
be larger than at any point outside.  In the one-dimensional case, a credible interval thus constructed 
corresponds to the Minimal Credible Interval of~\cite{Hamann:2007pi}. Provided the data are sufficiently constraining the marginal posterior typically exhibits very little dependence on
the choice of prior. For data that can only provide an upper or a lower bound on
a parameter (or no bound at all) however, the properties of the
inferred posterior and the boundaries of credible regions can vary
significantly with the choice of prior as well as its limits $\theta_{\rm min}$
and $\theta_{\rm max}$, making an objective interpretation of the
results rather difficult. This is in particular the case for the inference of credible regions subject to direct detection data. 

Indeed in the case of Xe100, CDMS and CRESST we construct intervals on the volume of the marginal posterior in $S$-space $\mathcal{P}_{\rm mar}(S|X)$,
where $S$ is the expected WIMP signal, using a uniform prior on $S$ with a lower boundary at 
zero. An $x$\% upper bound thus constructed has a well-defined Bayesian interpretation that 
the probability of $S\leq S_x$ is $x$\%.  The limit $S_x$ is then mapped onto the  \{$m_{\rm DM},\delta$\}-plane
by identifying those combinations of  $m_{\rm DM}$ and $\delta$ with $\mathcal{P}_{\rm mar}(m_{\rm DM},\delta|X) =\mathcal{P}_{\rm mar}(S_x|X)$.
An $x$\% contour, which will be denoted $x_S\%$, computed in this manner has the property of being independent of the choice of prior boundaries for $m_{\rm DM}$  and $\delta$. Its drawback, however, is that it has no well-defined probabilistic interpretation in \{$m_{\rm DM},\delta$\}-space.

\section{Standard Model Halo results for inelastic scattering}\label{app2}

\begin{figure}
\centering
\includegraphics[width=0.9\columnwidth]{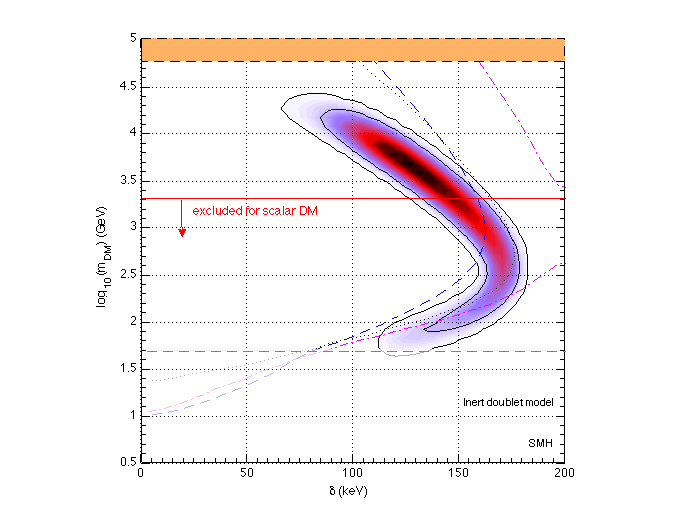}
\caption{2D credible regions for the individual experimental bounds and regions assuming the SMH, combined in a single plot. For DAMA (shaded) we show the 90\% and 99\% contours.
The $90_S\%$ contours are given by respectively the pink dot-dash curve for Xe100, the dashed blue line for CDMS and the dark green one denotes CRESST. The region below $M_\chi =$ 2 TeV is excluded by $\chi_0-\bar{\chi}_0$ oscillation (horizontal red solid line), while the orange/dark grey region is excluded by unitarity bound and below the dashed gray line by the LEP constraints on the Z decay width. The astrophysical observables are fixed at $\bar{v}_0$, $\bar{v}_{\rm esc}$ and $\bar{\rho}_\odot$. 
\label{fig:SallSMH}}
\end{figure}

For sake of completeness we show in figure~\ref{fig:SallSMH} the preferred region for DAMA modulation assuming the standard maxwellian halo with fixed astrophysical observables as DM velocity distribution, Eq.~\ref{eq:eta}. The astrophysical variables are fixed at their preferred values, see section~\ref{sec:intheo}. In the same plot we add the exclusion limits of CDMS, Xe100 and CRESST. The trend is analogous to figure~\ref{fig:Sall}, even though the DAMA region is smaller, since there a no volume effect due to marginalization over the astrophysical uncertainties. The region is also smaller because of $\bar{v}_{\rm esc}=544\  {\rm km s^{-1}}$, which limits the contribution of particles in the very high tail of the distribution. At the same time the exclusion bounds are tighter than in NFW and isothermal case, again because the reduce parameter space. The results regarding the viability of the DM candidate are practically unchanged with respect to the case with marginalization over the astrophysics: an asymmetric scalar DM with $M_\chi \gae 2 TeV$ as explanation of the DAMA signal is completely disfavoured by Xe100, while a fermionic component is compatible at $90\%$ C.L. up to a mass of 200 GeV.

\section*{References}
\bibliographystyle{iopart-num.bst}
\bibliography{biblio}

\end{document}